\UseRawInputEncoding
\documentclass[journal]{IEEEtran}
\usepackage{amsmath,amssymb,amsfonts}
\usepackage{subcaption}
\usepackage{algorithmic}
\usepackage{array}
\usepackage{stfloats}
\usepackage{tabularx}
\usepackage{textcomp}
\usepackage{xcolor}
\usepackage{url}

\usepackage{algorithm}
\usepackage{booktabs}
\usepackage{url}
\usepackage{makecell}
\usepackage{multirow}
\usepackage{verbatim}
\usepackage{graphicx}
\usepackage{cite}
\usepackage[switch]{lineno}
\usepackage{hyperref} 
\hypersetup{
    colorlinks=true, 
    citecolor=blue,
    linkcolor=blue,   
    filecolor=blue,     
    urlcolor=black       
}
\def\BibTeX{{\rm B\kern-.05em{\sc i\kern-.025em b}\kern-.08em
    T\kern-.1667em\lower.7ex\hbox{E}\kern-.125emX}}
\usepackage{balance}
\begin{document}
\title{A Generative AI-Enhanced Digital Twin Framework for Proactive Interference Management in Hybrid Near/Far-Field Wireless Systems}
\author{Afan Ali,~\IEEEmembership{Senior Member,~IEEE}, Ali Arshad Nasir,~\IEEEmembership{Senior Member,~IEEE}, and \\Daniel Benevides da Costa,~\IEEEmembership{Senior Member,~IEEE}
\thanks{A preliminary version of this paper has been presented in IEEE International Symposium on Personal, Indoor and Mobile Radio Communications (PIMRC), 1-4 September 2026, Singapore.}
\thanks{The authors are with the Interdisciplinary Research Center for Communication Systems and Sensing (IRC-CSS), Department of Electrical Engineering, King Fahd University of Petroleum and Minerals (KFUPM), Dhahran 31261, Saudi Arabia (e-mail: afan.ali@kfupm.edu.sa; anasir@kfupm.edu.sa; danielbcosta@ieee.org).}}

\markboth{Journal of \LaTeX\ Class Files,~Vol.~, No.~, May~2026}%
{How to Use the IEEEtran \LaTeX \ Templates}

\maketitle

\begin{abstract}
The applications of Digital Twins (DTs) and Generative AI (GenAI) have demonstrated significant potential for modeling and learning-based wireless communications. However, their joint potential for proactive wireless system design remains largely underexplored, particularly in extremely large-scale multiple-input multiple-output (XL-MIMO) networks characterized by hybrid near-field (NF) and far-field (FF) propagation. In this work, we propose an integrated GenAI-enhanced DT framework for proactive interference management in dynamic indoor environments. The DT constructs a site-specific virtual replica of the deployment, enabling geometry-aware modeling of blockage and hybrid NF/FF propagation. Conditioned on the DT environment and historical network states, GenAI generates multiple plausible future channel and blockage realizations, which are incorporated into regime-aware proactive beamforming. Extensive Sionna ray-tracing simulations across open and cluttered indoor environments, including different user equipment (UE) speeds, demonstrate improved interference suppression, signal-to-interference-plus-noise ratio (SINR), and prediction accuracy compared with reactive and deterministic DT-based benchmarks.
\end{abstract}


\section{Introduction}

\IEEEPARstart{N}{ext}-generation wireless networks are expected to support ultra-dense indoor deployments with extreme key performance indicators (KPIs), including terabit-per-second data rates and sub-millisecond latency~\cite{shafie_terahertz_2023}. Millimeter-wave (mmWave) and terahertz (THz) operation can support these requirements, but suffers from severe path loss and blockage sensitivity~\cite{jornet_channel_2011}. Extremely large-scale multiple-input multiple-output (XL-MIMO) provides substantial spatial gains; however, its large aperture places many users in the near-field (NF), creating hybrid NF/far-field (FF) environments where spherical and planar wavefronts coexist~\cite{ma_hybrid_2025}. 
This coexistence makes interference management particularly challenging because NF interference depends jointly on angle and distance, requiring distance-aware focusing rather than conventional angular processing. Moreover, mobility can cause links to cross the Rayleigh boundary, while blockage further alters the propagation and interference conditions. Consequently, instantaneous-CSI-based approaches~\cite{parvini_interference_2026,kokkoniemi_stochastic_2021,spencer_zero-forcing_2004,hu_near-field_2022} may become ineffective when these conditions evolve before beamforming adaptation is completed, motivating proactive prediction of future channel, blockage, and NF/FF states.

Digital Twins (DTs) provide a promising means of managing such dynamics through site-specific virtual representations of wireless environments. By capturing XL-MIMO arrays, propagation geometry, and dynamic users, DTs enable environment-aware network monitoring and optimization~\cite{tao_wireless_2024}. Existing studies have investigated DTs for real-time network control~\cite{deng_digital_2023}, sustainable in-network computing~\cite{becattini_digital_2024}, O-RAN~\cite{al-tahmeesschi_enhancing_2025}, and proactive network intelligence~\cite{peng_guest_2025}. 
Nevertheless, conventional deterministic DT implementations primarily replicate or estimate network states and may not represent multiple plausible futures under uncertain mobility, blockage, and hybrid-field dynamics. This motivates augmenting the physical consistency of DTs with stochastic predictive capabilities.

Generative AI (GenAI) provides such capabilities by learning complex data distributions and synthesizing plausible network states. Generative adversarial networks (GANs), variational autoencoders (VAEs), diffusion models, and related architectures support data generation and predictive modeling~\cite{goodfellow_generative_2014,kingma_auto-encoding_2013,ho_denoising_2020,vaswani_attention_2017}. Recent wireless studies have explored GenAI for channel modeling, resource allocation, network optimization, and intelligent management~\cite{karapantelakis_generative_2024,vu_applications_2025,khoramnejad_generative_2025}. However, these approaches generally employ GenAI as an isolated learning component rather than integrating it with a continuously synchronized, site-specific physical--virtual representation.

Consequently, a key research gap remains in combining the physics-consistent environment awareness of DTs with GenAI-based stochastic prediction for proactive hybrid-field interference management. Such integration can jointly anticipate future channel evolution, blockage, mobility, and NF/FF transitions and use these predictions to configure interference-management decisions before adverse conditions occur.

\subsection{Related Works}

\subsubsection{The Need for Precise Digital Replication}
XL-MIMO requires hybrid-field modeling beyond conventional planar-wave assumptions. Existing studies have investigated NF beam focusing, spatial non-stationarity, and distance-dependent channel characteristics~\cite{vaquero_near-field_2026,lee_nonuniform_2023,qin_fast_2024}, as well as unified NF/FF models~\cite{ma_hybrid_2025,nasir_max-min_2024,arshad_nasir_max-min_2025,liu_near-field_2023,lin_terahertz_2016}. 
NF spherical-wave propagation introduces joint angle--distance dependence, enabling location-domain focusing while creating distance-dependent interference and mobility-driven NF/FF transitions.
However, existing models are largely analytical or scenario-specific and do not fully capture the joint evolution of mobility, environment, and interference. 
DTs address this limitation through a site-specific, geometry-aware representation incorporating three-dimensional (3D) deployment geometry, antenna locations, material properties, blockage, and propagation paths, thereby linking environmental changes to channel and interference evolution.

\subsubsection{From Static Mitigation to DT-Enabled Interference Management}
Building on such geometry-aware representations, DTs can extend interference management beyond conventional instantaneous beamforming and scheduling. Traditional mmWave/THz approaches largely assume FF propagation~\cite{chang_joint_2022,dabiri_pointing_2022}, while recent NF solutions incorporate spherical-wave effects but remain predominantly reactive~\cite{ma_hybrid_2025}. DT-assisted ray tracing enables site specific interference coordination and synthetic channel generation~\cite{tao_wireless_2024,guan_integrating_2025}. 

However, DT fidelity requires synchronization with changes in mobility, blockage, and the environment, while repeated high-fidelity ray tracing can introduce significant latency. This motivates a multi-timescale architecture with slow DT construction and fast measurement-driven prediction and control.

\begin{table}[t]
\caption{Comparison of the proposed GenAI-enhanced DT framework with existing works}
\label{table:comparison}
\centering
\renewcommand{\arraystretch}{1.2}

\begin{tabular}{|l|p{1.1cm}|p{1.2cm}|p{1.3cm}|p{1.5cm}|}
\hline
\textbf{Literature} & \textbf{Far-field (FF)} & \textbf{Near-field (NF)} & \textbf{Hybrid (FF+NF)} & \textbf{Interference suppression} \\
\hline

\cite{duran_gentwin_2025,chai_generative_2024, naeem_survey_2025,li_generative_2025, chiaro_generative_2025,basaran_gen-twin_2025,tao_wireless_2024,fu_generative_2026,duran_generative_2024,savaglio_generative_2025,singh_network_2025,huang_when_2025}
& $\checkmark$ & - & - & - \\
\hline

\cite{jin_gdm4mmimo_2026}
& $\checkmark$ & $\checkmark$ & - & - \\
\hline

\cite{zhang_digital_2023,li_exploring_2026,kalor_wireless_2025}  
& $\checkmark$ & - & - & $\checkmark$ \\
\hline

\textbf{Our work} 
& $\checkmark$ & $\checkmark$ & $\checkmark$ & $\checkmark$ \\
\hline

\end{tabular}
\end{table}

\subsubsection{The Synergy of GenAI and DTs for Interference Management}

Beyond DT-based environment replication, GenAI provides stochastic modeling capabilities for anticipating uncertain network evolution. Recent works employ GANs and diffusion models for channel/RF data generation~\cite{chai_generative_2024,yang_large_2025,basaran_gen-twin_2025}, while GenAI-enabled DT architectures are surveyed in~\cite{li_generative_2025,naeem_survey_2025}. 

At the physical layer, generative diffusion models (GDMs) have been applied to NF XL-MIMO channel estimation~\cite{jin_near-field_2026} and conditional GDMs (CGDMs) to channel-fingerprint construction in MIMO systems~\cite{jin_channel_2026}. However, these studies focus primarily on estimation or fingerprinting rather than multi-step prediction for proactive interference management. Conditional GANs (cGANs) can instead generate multiple plausible futures conditioned on DT-derived environmental and historical information, while their single-pass inference provides a favorable latency tradeoff over iterative diffusion sampling for time-sensitive control.

Despite these advances, DTs, GenAI, and hybrid-field interference management have largely been studied separately. 
In particular, jointly predicting future channels, geometry-aware blockage, interference, and NF/FF transitions within a synchronized site-specific DT remains largely unexplored.
To the best of our knowledge, this is the first integrated GenAI-enhanced DT framework for proactive interference management in hybrid-field systems. 
It combines physics-consistent DT modeling, stochastic multi-trajectory prediction, and regime-aware proactive beamforming within a closed loop.
Table~\ref{table:comparison} summarizes the key distinctions from representative existing works.

\subsection{Motivation}

The discussion above highlights the need to integrate DT-based environment awareness with stochastic future-state prediction for hybrid-field interference management. 

In XL-MIMO systems, mobility, blockage, and NF/FF transitions jointly alter the channel and the appropriate beamforming strategy, making single-state or reactive solutions insufficient.

This motivates a framework that: (i) maintains a site-specific DT to capture geometry, blockage, and scatterers; (ii) generates multiple plausible future channel, blockage, and interference trajectories; and (iii) exploits the predicted NF/FF states for regime-aware proactive beamforming. Together, these capabilities combine the physical consistency of DTs with the stochastic foresight of GenAI to anticipate and mitigate future interference.

The following contributions summarize how the proposed framework realizes this integration.

\begin{figure*}[t!]
     \centering
     \begin{subfigure}[b]{0.3\textwidth}
         \centering
         \includegraphics[width=\textwidth]{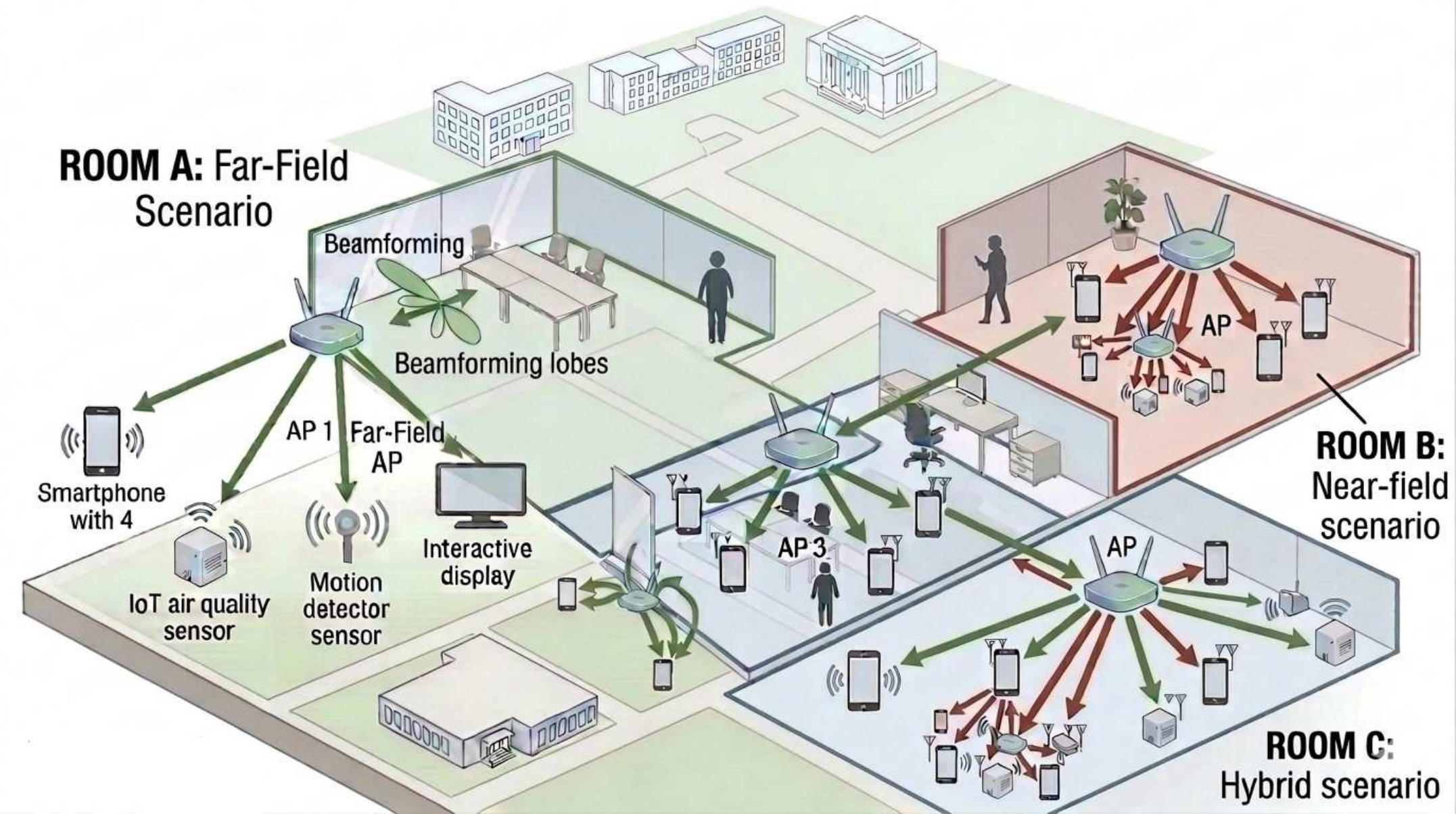}
         \vspace{-1.5ex}
         \caption{An illustration of dense indoor wireless communication environment.}
         
         \label{fig:inter_dense}
     \end{subfigure}
     \hfill
     \begin{subfigure}[b]{0.3\textwidth}
         \centering
         \includegraphics[width=\textwidth]{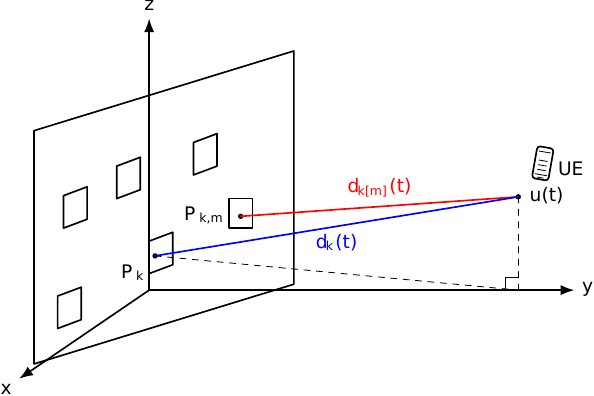}
         \vspace{-2ex}
         \caption{XL-MIMO UPA--UE link geometry.}
         
         \label{fig:system_model}
     \end{subfigure}
     \hfill
     \begin{subfigure}[b]{0.3\textwidth}
         \centering
         \includegraphics[width=\textwidth]{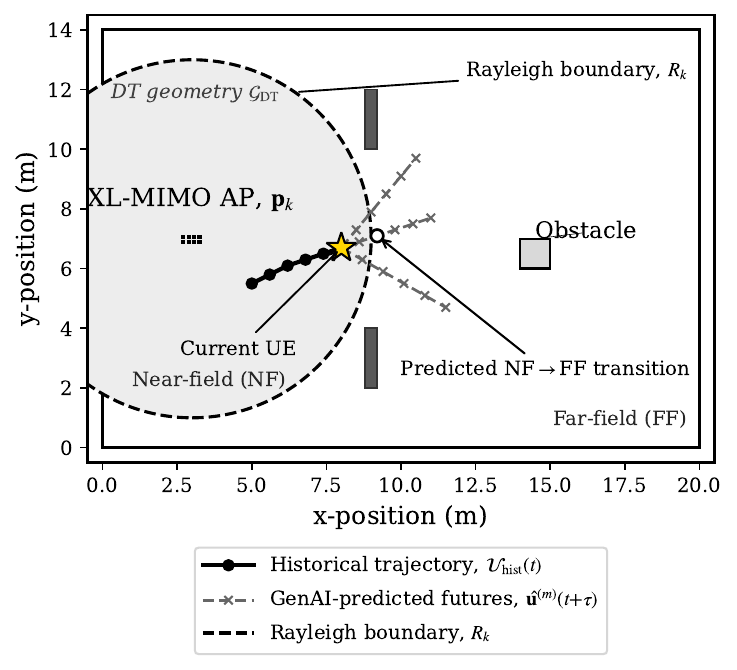}
         \vspace{-2ex}
         \caption{DT-assisted predictive mobility and NF/FF regime transition.}
         
         \label{fig:1c}
     \end{subfigure}
        \caption{Overall System Model illustration.}
        \vspace{-2ex}
        \label{fig:over_allsystem_diagram}
\end{figure*}

\subsection{Contributions}

The main contributions of this work are summarized as follows:

\begin{itemize}

\item

We propose an integrated GenAI-enhanced DT framework for stochastic future-state prediction and proactive interference management in hybrid NF/FF XL-MIMO systems. The framework explicitly incorporates spherical-wave propagation, product-distance multipath modeling, location-aware NF focusing, and regime-dependent precoding, enabling adaptation as links cross the Rayleigh boundary.

\item

To the best of our knowledge, most existing approaches model blockage as an environment-independent stochastic event. We develop a geometry-aware predictive blockage mechanism that combines DT-side 3D geometry with GenAI-predicted UE mobility to anticipate blockage and line of sight (LoS)/non-line of sight (NLoS) transitions and incorporate them into future effective channels.

\item

We develop a closed-loop proactive optimization framework that performs regime-aware robust beamforming over multiple GenAI-generated future channel, blockage, and NF/FF realizations rather than a single prediction. Physical-network feedback further synchronizes the DT--GenAI pipeline with the evolving environment.

\item

We validate the framework using Sionna ray-tracing in two distinct indoor environments, including NF/FF coexistence, different user equipment (UE) speeds, and generated-trajectory diversity. Comparisons with FF reactive ZF~\cite{spencer_zero-forcing_2004}, reactive hybrid-field beamforming~\cite{hu_near-field_2022}, deterministic DT prediction~\cite{khan_digital_2026}, and regime-unaware GenAI-DT prediction~\cite{li_vehicle_2021} demonstrate gains in interference, signal-to-interference-plus-noise ratio (SINR), outage, and prediction root-mean-square error (RMSE). We further evaluate floating point operations per seconds (FLOPs), model complexity, and online latency.

\end{itemize}
The remainder of this paper is organized as follows. Section~\ref{Sec:II} presents the system model. Section~\ref{Sec:III} formulates the problem 
and introduces the preliminary solution framework. 
Section~\ref{Sec:IV} details the proposed GenAI-enhanced DT framework. Section~\ref{Sec:V} 
presents the simulation setup, performance evaluation, and corresponding discussions, 
and Section~\ref{conclusion} concludes the paper.

\section{System Model}
\label{Sec:II}

We consider a high-fidelity DT framework for a dense indoor wireless environment, as illustrated in Fig.~\ref{fig:inter_dense}. The DT maintains a real-time virtual replica of the physical world, capturing site-specific geometries and electromagnetic (EM) properties to provide a structured training ground for GenAI. 
Specifically, the DT incorporates the 3D deployment geometry, transmitter and antenna-element locations, environmental objects, and their associated EM properties, thereby providing the physical context required for site-specific channel synthesis and predictive network modeling.
Fig.~\ref{fig:system_model} depicts a zoomed-in illustration of a specific link between an XL-MIMO uniform planar array (UPA) and a UE. 
Furthermore, Fig.~\ref{fig:1c} illustrates the temporal evolution of the considered hybrid-field scenario, where the DT combines the historical UE trajectory with the site-specific geometry to support GenAI-based prediction of future UE states and anticipate transitions across the NF/FF boundary.
The DT is synchronized with the physical environment using updated network and mobility information, while the detailed construction, calibration, and online update mechanisms of the site-specific DT are described in Section~\ref{Sec:IV}.
Table~\ref{tab:notation} summarizes the main notation used throughout this paper.

\subsection{DT-Assisted Topology and Predictive Mobility}

The DT tracks a set of transmitters $\mathcal{K}=\{0,1,\ldots,K\}$, where $k=0$ is the serving access point (AP), $\mathcal{U}$ represents 
the UE service region,
and $\mathcal{K}_I=\{1,\ldots,K\}$ represents the 
set of interfering transmitters.
By integrating a site-specific ray-tracing engine, the DT maintains the UE's trajectory, ${\bf u}(t) \in \mathbb{R}^3$, and determines the distance to each transmitter $k$ at position ${\bf p}_k$ as
\begin{equation}
d_k({\bf u}(t)) = \|{\bf u}(t) - {\bf p}_k\|_2,
\label{eq:ue_distance}
\end{equation}
where the UE position is continuously associated with the site-specific geometry represented by the DT. Consequently, the mobility state is interpreted relative to physical structures such as walls, obstacles, and transmitter locations rather than independently of the deployment environment.

Unlike traditional stochastic models, the GenAI module leverages the DT's 3D environmental map to predict the future state ${\bf u}(t+\tau)$. 
Specifically, the GenAI module processes a sequence of $N_h$ historical UE position samples, defined as
\begin{equation}
\mathcal{U}_{\mathrm{hist}}(t)
=
\left\{
{\bf u}\big(t-(N_h-1)\Delta t\big),\ldots,
{\bf u}(t-\Delta t),{\bf u}(t)
\right\},
\label{eq:mobility_history}
\end{equation}
where $N_h$ denotes the number of historical UE position samples and $\Delta t$ is the sampling interval.
By processing this historical mobility information together with the site-specific environmental knowledge maintained by the DT, the GenAI anticipates how the UE will evolve relative to physical boundaries, enabling the network to prepare for regime transitions before they occur in the physical layer. 
As conceptually illustrated in Fig.~\ref{fig:1c}, multiple plausible future mobility realizations can evolve from the same historical trajectory due to uncertainty in user motion. The DT geometry constrains these predictions relative to environmental obstacles and the Rayleigh boundary, thereby providing the spatial context required to anticipate future blockage conditions and NF/FF regime transitions.

\subsection{GenAI-Enhanced Hybrid-Field Modeling}

The extremely large physical aperture of XL-MIMO arrays results in a significant Rayleigh distance defined as
\begin{equation}
R_k
=
\frac{2D_k^2}{\lambda},
\label{eq:rayleigh_distance}
\end{equation}
where $\lambda$ denotes the carrier wavelength and $D_k$ is the 
physical aperture
of transmitter $k$. 
The Rayleigh distance provides the criterion used to distinguish the NF region, characterized by spherical-wave propagation and distance-dependent spatial focusing, from the FF region, where the planar-wave approximation is applicable. Owing to the large aperture of XL-MIMO arrays, NF and FF links may coexist within the same deployment, thereby requiring propagation-regime-aware channel modeling and beamforming.

Accordingly, the DT dynamically classifies each link into NF or FF regimes as
\begin{equation}
k \in
\begin{cases}
\mathcal{K}_{\mathrm{N}}(\mathbf{u}(t)), & d_k(\mathbf{u}(t)) \le R_k, \\
\mathcal{K}_{\mathrm{F}}(\mathbf{u}(t)), & d_k(\mathbf{u}(t)) > R_k,
\end{cases}
\label{eq:regime_classification}
\end{equation}
where $\mathcal{K}_{\mathrm{N}}(\mathbf{u}(t))$ and
$\mathcal{K}_{\mathrm{F}}(\mathbf{u}(t))$ denote the sets of NF and FF
transmitters, respectively.
This classification is dynamically updated as the UE position evolves. As illustrated in Fig.~\ref{fig:1c}, a UE currently located in the NF may have predicted future positions beyond the Rayleigh boundary. For proactive operation, the same criterion in equation \eqref{eq:regime_classification} is therefore evaluated at the GenAI-predicted position ${\bf u}(t+\tau)$, allowing the framework to anticipate NF-to-FF or FF-to-NF transitions before they occur. The resulting regime information is subsequently used by the hybrid-field channel model and regime-aware beamforming strategy.

\begin{table}[t]
\caption{Summary of Main Notation}
\label{tab:notation}
\centering
\scriptsize
\renewcommand{\arraystretch}{0.95}
\setlength{\tabcolsep}{2.5pt}
\begin{tabular}{|l|p{5.0cm}|}
\hline
\textbf{Symbol} & \textbf{Description} \\
\hline

$\mathcal{K}$, $\mathcal{K}_I$
& All and interfering transmitters \\
\hline

$\mathcal{K}_{\mathrm N}$, $\mathcal{K}_{\mathrm F}$
& NF and FF transmitter sets \\
\hline

$M_k$, $L$
& Antenna elements and propagation paths \\
\hline

$\mathbf{u}(t)$, $\mathbf{p}_k$, $\mathbf{p}_{k,m}$
& UE, transmitter, and antenna-element positions \\
\hline

$\tilde{\mathbf p}_{k,l}$
& Position of the $l$-th scatterer \\
\hline

$d_k$, $d_{k,l}[m]$
& AP--UE and element-wise path distances \\
\hline

$D_k$, $R_k$, $\lambda$
& Array aperture, Rayleigh distance, and wavelength \\
\hline

$\mathbf h_k$, $\mathbf h_{k,\mathrm{DT}}$
& Physical and DT-synthesized channels \\
\hline

$\hat{\mathbf h}_k$, $\hat{\mathbf h}^{\mathrm{eff}}_k$
& Predicted and effective predicted channels \\
\hline

$\mathbf h_{\mathrm{BS}}$, $h_{\mathrm{UE}}$
& BS- and UE-side spherical responses \\
\hline

$\beta_k$, $\Lambda_k$
& Large-scale fading and path loss \\
\hline

$\xi_k$, $\hat{\xi}_k$, $\eta$
& Blockage factors and severe-blockage attenuation \\
\hline

$p_k^{\mathrm{blk}}$
& Predicted blockage probability \\
\hline

$\rho_k$
& NF/FF regime indicator \\
\hline

$\mathbf w_k$, $\mathbf w_k^{\mathrm{pro}}$
& Beamforming and proactive beamforming vectors \\
\hline

$P_k$, $P_{\mathrm{total}}$, $\sigma^2$
& Transmit power, power budget, and noise power \\
\hline

$I$, $\hat I$, $I_{\mathrm{DT}}$
& Interference: measured, predicted, and DT-generated \\
\hline

$\gamma$, $\hat{\gamma}$, $\gamma_{\min}$
& SINR, predicted SINR, and minimum SINR \\
\hline

$G_{\mathrm{DT}}$, $\mathcal U_{\mathrm{hist}}$
& DT environment and historical UE trajectory \\
\hline

$\mathbf X_d^{\mathrm{DT}}$, $\mathbf Y_d^{\mathrm{DT}}$
& DT input--output training samples \\
\hline

$c(t)$, $\mathbf z$, $G(\cdot)$, $D(\cdot)$
& cGAN context, latent variable, generator, and discriminator \\
\hline

$\hat{\mathcal Y}(t)$, $M$
& Generated future trajectories and number of samples \\
\hline

$T$, $\Delta t$, $\tau$
& Prediction horizon, time step, and prediction offset \\
\hline

$\mathcal L_{\mathrm{adv}}$, $\mathcal L_{\mathrm{pred}}$
& Adversarial and prediction losses \\
\hline

\end{tabular}
\end{table}
\subsubsection{DT-Driven Predictive Blockage and Path Loss}

In the considered dense indoor environment, the large-scale fading coefficient of the $k$-th link can be modeled as
\begin{equation}
\beta_k(\mathbf{u}(t),t)
=
\xi_k(t)\Lambda_k(\mathbf{u}(t),t),
\label{eq:large_scale_fading_refined}
\end{equation}
where $\xi_k(t)\in\{1,\eta\}$ denotes the blockage indicator, $\eta\ll1$ represents the attenuation factor during severe blockage, and $\Lambda_k(\mathbf{u}(t),t)$ denotes the large-scale path-loss component. The distance-dependent path loss can be expressed as
\begin{equation}
\Lambda_k(\mathbf{u}(t),t)
=
C_0
\left(
\frac{d_k(\mathbf{u}(t))}{d_{\mathrm{ref}}}
\right)^{-\alpha},
\label{eq:path_loss_refined}
\end{equation}
where $C_0$ is the reference path-loss constant, $d_{\mathrm{ref}}$ is the reference distance, and $\alpha$ denotes the path-loss exponent.

Conventional wireless models typically treat the blockage process $\xi_k(t)$ as a random stochastic variable. In contrast, the proposed DT-assisted GenAI framework transforms blockage into a geometry-aware predictive quantity.

Specifically, the DT obtains the instantaneous blockage state from the site-specific propagation geometry. For each transmitter--UE link, the LoS and dominant propagation paths provided by the ray-tracing engine are evaluated with respect to the walls, obstacles, and other blocking objects represented in $G_{\mathrm{DT}}$. A link is assigned $\xi_k(t)=1$ when the corresponding propagation path is unobstructed, whereas $\xi_k(t)=\eta$ when severe blockage is detected. Similarly, the locations of dominant scatterers $\tilde{\mathbf{p}}_{k,l}$ are obtained from the propagation paths identified by the site-specific ray-tracing engine. Therefore, the blockage state and scatterer information are derived from the DT environment rather than being prescribed independently of the physical deployment.

Since the DT maintains the 3D geometry and associated electromagnetic (EM) properties of the indoor environment, the GenAI module can infer future blockage events by analyzing the predicted interaction between the UE trajectory and surrounding obstacles.

The predicted UE motion is evaluated relative to the DT-side obstacles and propagation geometry, enabling the framework to anticipate potential obstruction of the transmitter--UE link before the UE reaches the corresponding location.

Consequently, the predicted future channel at time $t+\tau$ can be expressed as
\begin{equation}
\hat{\mathbf{h}}_k(t+\tau)
=
\sqrt{\hat{\beta}_k(t+\tau)}
\hat{\mathbf{g}}_k(t+\tau),
\label{eq:predicted_channel_refined}
\end{equation}
in which $\hat{\mathbf{g}}_k(t+\tau)$ denotes the predicted small-scale hybrid-field channel and $\hat{\beta}_k(t+\tau)$ denotes the predicted large-scale fading coefficient. The predicted large-scale fading coefficient can be given as
\begin{equation}
\hat{\beta}_k(t+\tau)
=
\hat{\xi}_k(t+\tau)
\hat{\Lambda}_k(\mathbf{u}(t+\tau),t+\tau),
\label{eq:predicted_beta_refined}
\end{equation}
in which $\hat{\xi}_k(t+\tau)$ is the predicted blockage indicator and $\hat{\Lambda}_k(\mathbf{u}(t+\tau),t+\tau)$ is the predicted path loss.

Rather than predicting the blockage state independently of the remaining propagation conditions, the proposed framework jointly predicts

\begin{equation}
\left(
\hat{\mathbf{h}}_k(t+\tau),
\hat{\Lambda}_k(t+\tau),
\hat{\xi}_k(t+\tau)
\right),
\label{eq:joint_prediction}
\end{equation}
thereby enabling geometry-aware prediction of future propagation conditions.

The GenAI mapping can be formulated as
\begin{equation}
f_{\mathrm{GenAI}}:
(\mathcal{U}_{\mathrm{hist}},G_{\mathrm{DT}})
\mapsto
\left(
\hat{\mathbf{h}}_k(t+\tau),
\hat{\Lambda}_k(t+\tau),
\hat{\xi}_k(t+\tau)
\right),
\label{eq:genai_mapping_refined}
\end{equation}
where $G_{\mathrm{DT}}$ denotes the DT-side environment representation containing the 3D geometry and EM characteristics of the physical environment, and $\mathcal{U}_{\mathrm{hist}}$ denotes the historical mobility trajectory.

Since $\xi_k(t)\in\{1,\eta\}$ is a discrete blockage factor, its conditional expectation generally takes a continuous value between $\eta$ and $1$ and therefore does not itself constitute a physically valid realization of the blockage state. Accordingly, the GenAI module instead estimates the conditional probability of future blockage as

\begin{equation}
p_k^{\mathrm{blk}}(t+\tau)
=
\Pr\!\left(
\xi_k(t+\tau)=\eta
\,\middle|\,
\mathcal{U}_{\mathrm{hist}},G_{\mathrm{DT}}
\right).
\label{eq:blockage_prob}
\end{equation}

Here, $p_k^{\mathrm{blk}}(t+\tau)\in[0,1]$ denotes the predicted probability that the $k$-th link will experience severe blockage at time $t+\tau$, conditioned on the historical UE trajectory $\mathcal{U}_{\mathrm{hist}}$ and the site-specific environmental information contained in $G_{\mathrm{DT}}$. Accordingly, larger values of $p_k^{\mathrm{blk}}(t+\tau)$ indicate a higher likelihood that the future blockage factor will take the attenuated state $\xi_k(t+\tau)=\eta$. For stochastic future-state generation, $\hat{\xi}_k(t+\tau)$ is sampled according to this predicted blockage probability, whereas a binary decision based on the same probability can be employed when a single representative future state is required. In either case, the resulting predicted blockage factor remains restricted to $\hat{\xi}_k(t+\tau)\in\{1,\eta\}$, thereby preserving consistency with the physical blockage model in equation \eqref{eq:large_scale_fading_refined}.

In contrast to the discrete blockage factor, the path-loss component $\Lambda_k(t+\tau)$ is a continuous quantity. Therefore, its conditional-mean interpretation remains appropriate under a mean-squared prediction criterion, yielding

\begin{equation}
\hat{\Lambda}_k(t+\tau)
\approx
\mathbb{E}
\left[
\Lambda_k(t+\tau)
\mid
\mathcal{U}_{\mathrm{hist}},
G_{\mathrm{DT}}
\right].
\label{eq:pathloss_expectation}
\end{equation}

Equations~\eqref{eq:blockage_prob} and \eqref{eq:pathloss_expectation} assume that the DT geometry remains sufficiently stable over the short-term prediction horizon. Static structures are treated as unchanged, while UE mobility and dynamic blockage are captured through updated states. As the prediction horizon increases, accumulated mobility and environmental uncertainty may reduce prediction accuracy, as evaluated in Section~\ref{Sec:V}.

\subsubsection{Near-Field Multi-Path Synthesis}

For users in the NF regime ($k\in\mathcal{K}_{\mathrm{N}}$), the channel incorporates spherical-wave propagation and spatial non-stationarity~\cite{liu_near-field_2023}. 
Unlike the conventional FF model, where the propagation distance across the array is approximated using a common planar-wave direction, the NF model retains the exact element-wise propagation distances. This allows the channel representation to capture both angular and distance-dependent spatial characteristics of XL-MIMO propagation.
The DT provides the coordinates of scatterers, $\tilde{\mathbf{p}}_{k,l}$, obtained via ray tracing.
Specifically, the site-specific ray-tracing engine identifies the dominant LoS and reflected propagation paths from the 3D environment represented by $G_{\mathrm{DT}}$, from which the corresponding scatterer locations and element-wise path distances are extracted.
The NF channel vector for $M_k$ antenna elements can be expressed as
\begin{align}
\mathbf{h}_k(\mathbf{u}(t),t)
&=
\sum_{l=0}^{L-1}
\sqrt{\beta^{\mathrm{(PD)}}_{k,l}(\mathbf{u}(t))}
\nonumber \\
&\quad \times
\left[
 e^{-j\frac{2\pi}{\lambda}d_{k,l}[1](t)},
 \ldots,
 e^{-j\frac{2\pi}{\lambda}d_{k,l}[M_k](t)}
\right]^T,
\label{eq:nf_channel_refined}
\end{align}
where $l=0$ corresponds to the LoS component, $l\ge1$ corresponds to the NLoS components, and $d_{k,l}[m](t)$ denotes the exact propagation distance to the $m$-th antenna element.
Hence, the phase progression across the XL-MIMO aperture is determined by the actual propagation distance associated with each antenna element rather than by a common FF steering direction. This element-wise representation is the principal NF characteristic exploited by the subsequent location-aware channel synthesis and beamforming stages.

Unlike conventional planar-wave models, the considered NF multipath channel explicitly follows the product-distance propagation law described in~\cite{liu_near-field_2023}. Therefore, the total NF channel can be decomposed into LoS and NLoS components as
\begin{equation}
\mathbf{h}_k(\mathbf{u}(t),t)
=
\mathbf{h}_k^{\mathrm{LoS}}(t)
+
\mathbf{h}_k^{\mathrm{NLoS}}(t),
\label{eq:los_nlos_decomposition}
\end{equation}
where the two components are determined by the corresponding geometry-resolved propagation paths.
The LoS component is given by
\begin{equation}
\mathbf{h}_k^{\mathrm{LoS}}(t)
=
\sqrt{\beta_{k,0}^{\mathrm{LoS}}(t)}
\left[
e^{-j\frac{2\pi}{\lambda}d_{k,0}[1](t)},
\ldots,
e^{-j\frac{2\pi}{\lambda}d_{k,0}[M_k](t)}
\right]^T,
\label{eq:los_component}
\end{equation}
where
\begin{equation}
\beta_{k,0}^{\mathrm{LoS}}(t)
=
\left(
\frac{\lambda}{4\pi d_k(\mathbf{u}(t))}
\right)^2,
\label{eq:los_path_gain}
\end{equation}
denotes the free-space LoS path gain.

For the NLoS component, each reflected path is modeled as a cascaded BS-to-scatterer-to-UE propagation process as~\cite{liu_near-field_2023}
\begin{equation}
\mathbf{h}_{k,l}^{\mathrm{NLoS}}(t)
=
\alpha_{k,l}(t)
\mathbf{h}_{\mathrm{BS}}
\left(r_{k,l}^{\mathrm{BS}}(t)\right)
h_{\mathrm{UE}}
\left(r_{k,l}^{\mathrm{UE}}(t)\right),
\quad l\ge 1.
\label{eq:nlos_product_distance}
\end{equation}
where $\alpha_{k,l}(t)$ is the complex reflection coefficient, $r_{k,l}^{\mathrm{BS}}(t)$ is the BS-to-scatterer distance, and $r_{k,l}^{\mathrm{UE}}(t)$ is the scatterer-to-UE distance. 
Here, $\mathbf{h}_{\mathrm{BS}}(\cdot)$ and $h_{\mathrm{UE}}(\cdot)$ denote the BS-side and UE-side spherical-wave propagation responses, respectively. In particular, $\mathbf{h}_{\mathrm{BS}}(r)\in\mathbb{C}^{M_k}$ represents the element-dependent propagation response between the XL-MIMO array and the corresponding scatterer, including the spherical-wave phase variation across the array, whereas $h_{\mathrm{UE}}(r)$ represents the scalar propagation response from that scatterer to the single-antenna UE. This decomposition explicitly preserves the two geometric segments of each reflected path instead of absorbing them into a single effective path coefficient.
Consequently, the NLoS attenuation can be directly related to the BS--scatterer and scatterer--UE propagation distances, preserving the geometry dependence required by the product-distance NF model.

The total NLoS channel can be written as
\begin{equation}
\mathbf{h}_k^{\mathrm{NLoS}}(t)
=
\sum_{l=1}^{L-1}
\mathbf{h}_{k,l}^{\mathrm{NLoS}}(t),
\label{eq:total_nlos_channel}
\end{equation}
The path-wise NLoS attenuation can be defined as
\begin{equation}
\Lambda_{k,l}^{\mathrm{NLoS}}(t)
=
\frac{
\lambda^2
|\alpha_{k,l}(t)|^2
}{
(4\pi)^2
r_{k,l}^{\mathrm{BS}}(t)
r_{k,l}^{\mathrm{UE}}(t)
},
\label{eq:nlos_path_loss}
\end{equation}
Accordingly, the total NLoS path-loss contribution can be written as
\begin{equation}
\Lambda_k^{\mathrm{NLoS}}(t)
=
\sum_{l=1}^{L-1}
\Lambda_{k,l}^{\mathrm{NLoS}}(t),
\label{eq:total_nlos_path_loss}
\end{equation}
and the total large-scale path-loss contribution becomes
\begin{equation}
\Lambda_k(t)
=
\beta_{k,0}^{\mathrm{LoS}}(t)
+
\Lambda_k^{\mathrm{NLoS}}(t),
\label{eq:total_path_loss_los_nlos}
\end{equation}
The LoS path gain is represented using the conventional free-space attenuation, whereas the NLoS components follow the cascaded product-distance formulation in equation~\eqref{eq:nlos_product_distance}.
Together, these expressions ensure that the NF channel depends explicitly on the UE location, antenna-element positions, and DT-derived scatterer geometry, rather than being obtained from a conventional FF angular approximation.

Moreover, recent GenAI-enabled XL-MIMO channel estimation works demonstrate that generative models can effectively learn high-dimensional NF channel distributions while preserving geometry-aware propagation characteristics~\cite{jin_gdm4mmimo_2026}. 
In the proposed framework, however, GenAI does not replace the underlying physics-based channel model. Instead, it predicts the future geometry-dependent propagation state, while the DT constrains the resulting channel synthesis according to the spherical-wave and product-distance relationships described above. This physics-guided construction prevents the generated future channels from being treated as unconstrained data-driven realizations.

In the proactive DT-GenAI framework, the future NF channel can be predicted at time $t+\tau$ as
\begin{align}
\hat{\mathbf{h}}_k(t+\tau)
&=
\sum_{l=0}^{L-1}
\sqrt{\hat{\beta}^{\mathrm{(PD)}}_{k,l}(t+\tau)}
\nonumber \\
&\quad \times
\left[
 e^{-j\frac{2\pi}{\lambda}\hat{d}_{k,l}[1](t+\tau)},
 \ldots,
 e^{-j\frac{2\pi}{\lambda}\hat{d}_{k,l}[M_k](t+\tau)}
\right]^T,
\label{eq:predicted_nf_channel_refined}
\end{align}
where $\hat{d}_{k,l}[m](t+\tau)$ denotes the predicted path distance and $\hat{\beta}^{\mathrm{(PD)}}_{k,l}(t+\tau)$ denotes the predicted product-distance path gain.
These quantities are obtained from the predicted UE state together with the DT-side antenna and scatterer geometry and the corresponding product-distance propagation model. Hence, even for a GenAI-predicted future state, the spatial structure of the NF channel remains explicitly constrained by the underlying physical geometry.

Thus, the predicted effective channel can be defined as
\begin{equation}
\hat{\mathbf{h}}_k^{\mathrm{eff}}(t+\tau)
=
\sqrt{
\hat{\Lambda}_k(t+\tau)
\hat{\xi}_k(t+\tau)
}
\hat{\mathbf{h}}_k(t+\tau),
\label{eq:effective_channel_refined}
\end{equation}
where $\hat{\mathbf{h}}_k(t+\tau)$ captures the NF spatial structure, $\hat{\Lambda}_k(t+\tau)$ captures path loss, 
and
$\hat{\xi}_k(t+\tau)$ captures blockage attenuation.
Accordingly, $\hat{\mathbf{h}}_k^{\mathrm{eff}}(t+\tau)$ combines the predicted NF spatial response with the geometry-aware path-loss and blockage states introduced in the preceding subsection.
The effective channel in equation~\eqref{eq:effective_channel_refined} is subsequently used for interference evaluation and beamforming optimization.

\subsection{Proactive Signal and Interference Model}

The received signal at the UE can be expressed as
\begin{equation}
y(t)
=
\underbrace{
\mathbf{h}_0^H(\mathbf{u}(t),t)
\mathbf{w}_0 s_0(t)
}_{\text{Desired signal}}
+
\underbrace{
\sum_{k\in\mathcal{K}_I}
\mathbf{h}_k^H(\mathbf{u}(t),t)
\mathbf{w}_k s_k(t)
}_{\text{Interference}}
+
\underbrace{n(t)}_{\text{Noise}},
\label{eq:received_signal_refined}
\end{equation}
where $\mathbf{w}_k$ denotes the beamforming vector, $s_k(t)$ denotes the transmitted signal, and $n(t)\sim\mathcal{CN}(0,\sigma^2)$ is additive white Gaussian noise (AWGN).

Considering a pedestrian UE with speed $v\approx1$ m/s and carrier frequency $f_c=100$ GHz, the maximum Doppler shift is
\begin{equation}
f_d
=
\frac{vf_c}{c}
\approx
333.3~\mathrm{Hz},
\end{equation}
which yields a coherence time of approximately
\begin{equation}
T_c
\approx
\frac{0.423}{f_d}
\approx
1.27~\mathrm{ms}.
\end{equation}

The beamforming update interval is selected to be comparable to or shorter than the coherence time, while the DT--GenAI framework predicts effective channels over a multi-step horizon. Thus, prediction may span multiple coherence intervals, with each beamforming update based on the corresponding short-term predicted channel state.

The predicted aggregate interference at time $t+\tau$ is expressed using the effective channel as
\begin{equation}
\hat{I}(t+\tau)
=
\sum_{k\in\mathcal{K}_I}
P_k
\left|
\left(
\hat{\mathbf{h}}_k^{\mathrm{eff}}(t+\tau)
\right)^H
\mathbf{w}_k
\right|^2,
\label{eq:predicted_interference_refined}
\end{equation}

where $\hat{\mathbf{h}}_k^{\mathrm{eff}}(t+\tau)$ jointly incorporates the predicted hybrid-field spatial response, path loss, and blockage attenuation.

The proactive SINR at time $t+\tau$ is then given by
\begin{equation}
\hat{\gamma}_0(t+\tau)
=
\frac{
P_0
\left|
\left(
\hat{\mathbf{h}}_0^{\mathrm{eff}}(t+\tau)
\right)^H
\mathbf{w}_0
\right|^2
}{
\sum_{k\in\mathcal{K}_I}
P_k
\left|
\left(
\hat{\mathbf{h}}_k^{\mathrm{eff}}(t+\tau)
\right)^H
\mathbf{w}_k
\right|^2
+
\sigma^2
},
\label{eq:proactive_sinr_refined}
\end{equation}

which provides the predicted link-quality metric used for proactive beamforming optimization. In practice, measured CSI and interference estimates are fed back to the DT for synchronization and prediction refinement.

\section{Problem Formulation}
\label{Sec:III}

To capture the fast temporal dynamics of indoor channels, we adopt a millisecond-scale prediction resolution. Specifically, the time step is set to $\Delta t=1$ ms, which is comparable to the coherence time at $100$ GHz under pedestrian mobility. Therefore, the proposed DT--GenAI framework targets short-term channel and interference prediction over multiple coherence intervals, enabling proactive beam adaptation before channel aging degrades the SINR. 
At each prediction step, the objective is to determine the beamforming vectors $\{\mathbf{w}_k\}$ using the predicted effective channel state at $t+\tau$, rather than the instantaneous channel available at time $t$.
By leveraging the GenAI-predicted channel states, blockage indicators, and large-scale fading, we formulate the proactive beamforming optimization problem as

\begin{subequations}
\begin{align}
\max_{\{\mathbf{w}_k\}} \quad 
& \hat{\gamma}_0(t+\tau) = 
\frac{
P_0 
\left|
\left(\hat{\mathbf{h}}^{\mathrm{eff}}_0(t+\tau)\right)^H
\mathbf{w}_0
\right|^2
}{
\sum_{k\in\mathcal{K}_I}
P_k
\left|
\left(\hat{\mathbf{h}}^{\mathrm{eff}}_k(t+\tau)\right)^H
\mathbf{w}_k
\right|^2
+\sigma^2
},
\label{eq:proactive_opt1}\\
\text{s.t.}\quad
& \sum_{k\in\mathcal{K}}
P_k\|\mathbf{w}_k\|^2
\le P_{\mathrm{total}},
\label{eq:power_budget}\\
& \hat{\xi}_k(t+\tau)\in\{1,\eta\},
\quad \forall k\in\mathcal{K}.
\label{eq:regime_aware}
\end{align}
\end{subequations}
where $P_{\mathrm{total}}$ is the network's total power budget.

The constraint in equation \eqref{eq:regime_aware} preserves consistency with the discrete blockage model, while the predicted blockage attenuation is already incorporated into $\hat{\mathbf{h}}^{\mathrm{eff}}_k(t+\tau)$. The NF/FF regime awareness is determined separately through the Rayleigh-distance criterion in equation \eqref{eq:regime_classification}, evaluated at the predicted UE position.
By optimizing for $t+\tau$, the DT--GenAI engine 
adapts the beamforming vectors to the anticipated channel, blockage, and propagation-regime conditions, thereby mitigating interference before the corresponding network state occurs.

\begin{figure*}[!t]
\centering
\includegraphics[width=4.5in]{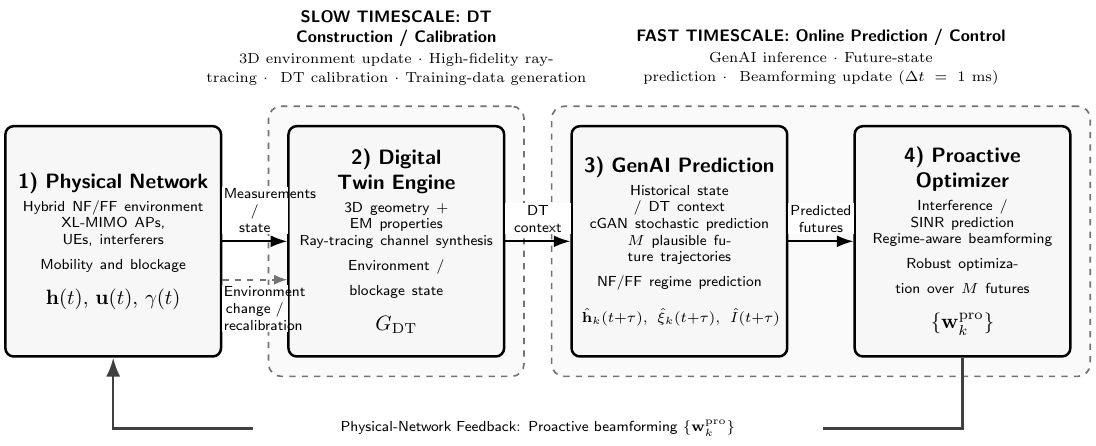}
\vspace{-1.5ex}
\caption{Closed-loop GenAI-enhanced DT architecture with slow-timescale DT construction/calibration and fast-timescale prediction, optimization, and physical-network feedback.}
\vspace{-2ex}
\label{BD-2}
\end{figure*}

\section{Proposed GenAI-Enhanced Digital Twin Framework}
\label{Sec:IV}

This section presents the integrated GenAI-enhanced DT framework designed for proactive interference management. 
The proposed framework combines site-specific physical modeling, stochastic future-state prediction, and regime-aware optimization within a closed-loop architecture.

\subsection{Overall Architecture}

The closed-loop architecture in Fig.~\ref{BD-2} comprises the physical network, site-specific DT, GenAI predictor, and proactive optimizer.
Physical-network observations synchronize the DT, whose geometry-aware context conditions the cGAN to generate multiple future channel,
blockage, and interference realizations. These predictions are supplied to the NF/FF-aware optimizer, and the resulting beamforming vectors are
deployed back to the physical network. DT construction/calibration operates at a slower timescale, whereas prediction and control constitute the fast online loop.

\subsection{Site-Specific DT-Aided Data Acquisition}
\label{subsec:dt_data}

The DT serves as a site-specific data acquisition and environment emulation engine ($G_{\mathrm{DT}}$) for the proposed framework~\cite{alkhateeb_real-time_2023, khan_digital_2026}. It constructs a high-fidelity virtual replica of the indoor deployment by modeling the physical environment, including transmitter locations $\{{\bf p}_k\}$, antenna-element positions $\{{\bf p}_{k,m}\}$, user trajectories ${\bf u}(t)$, and environmental objects such as walls and scatterers.
In practice, the static DT geometry can be initialized from available building information, such as floor plans, computer-aided design (CAD)/building information modeling (BIM) data, or 3D site scans, while transmitter and antenna locations are obtained from the deployment configuration. Material-dependent EM properties are assigned to the corresponding environmental surfaces and calibrated using available physical-network measurements. Dynamic quantities, including UE positions, mobility, blockage, and interference states, are subsequently updated from network observations. In the considered Sionna-based implementation, these quantities are instantiated directly from the site-specific 3D ray-tracing scenes and their associated material and propagation information.
This separation between relatively static geometric information and dynamically updated network states enables the DT to maintain environment fidelity without reconstructing the complete ray-traced environment at every online prediction step.

Let $\mathcal{E}(t)$ denote the physical environment at time $t$. The DT constructs a digital representation $\mathcal{E}_{\mathrm{DT}}(t)$ and generates channels through the propagation mapping
\begin{equation}
{\bf h}_{k,\mathrm{DT}}({\bf u}(t),t)
=
g_{\mathrm{DT}}\!\bigl(\mathcal{E}_{\mathrm{DT}}(t)\bigr),
\label{eq:dt_channel_gen1}
\end{equation}
where ${\bf h}_{k,\mathrm{DT}}({\bf u}(t),t)$ denotes the geometry-consistent channel synthesized by the DT via ray tracing and $g_{\mathrm{DT}}(\cdot)$ is the ray-tracing operator, consistent with the hybrid-field channel model in Section~\ref{Sec:II}.
The resulting propagation paths provide the channel coefficients, blockage states, and dominant scatterer information required by the environment-aware model introduced in Section~\ref{Sec:II}.

Using this model, the DT generates the training dataset
\begin{equation}
\mathcal{D}_{\mathrm{DT}}
=
\left\{
\left({\bf X}_d^{\mathrm{DT}}, {\bf Y}_d^{\mathrm{DT}}\right)
\right\}_{d=1}^{D_{\mathrm{DT}}},
\label{eq:dt_dataset_pairs}
\end{equation}
where ${\bf X}_d^{\mathrm{DT}}$ contains the historical, geometric, and environmental features and ${\bf Y}_d^{\mathrm{DT}}$ contains the corresponding future prediction targets.

Using the NF/FF classification defined in equation \eqref{eq:regime_classification}, each link is associated with the regime indicator $\rho_k(t)\in\{0,1\}$, where $\rho_k(t)=1$ and $\rho_k(t)=0$ denote NF and FF propagation, respectively.
The input feature vector is then defined as
\begin{equation}
{\bf X}_d^{\mathrm{DT}}
=
\Bigl[
{\bf H}_{t-T_h:t}^{\mathrm{hist}},
{\bf u}(t),
\{\rho_k(t)\}_{k\in\mathcal{K}},
{\bf r}_{\mathrm{event}}(t)
\Bigr],
\label{eq:dt_features}
\end{equation}
where ${\bf H}_{t-T_h:t}^{\mathrm{hist}}$ denotes the historical channel observations over a window of length $T_h$, and ${\bf r}_{\mathrm{event}}(t)$ encodes environment-level events such as blockage or hotspot formation.
Together, these features provide the GenAI module with temporal channel information, current mobility state, hybrid-field regime information, and environmental context.

The corresponding output consists of future channel and interference trajectories,
\begin{align}
{\bf Y}_d^{\mathrm{DT}} 
= 
\left\{
\left[
\left\{
\hat{\mathbf{h}}_{k}(t+\tau)
\right\}_{k\in\mathcal{K}},
\hat{I}(t+\tau)
\right]
\right\}_{\tau=1}^{T},
\label{eq:dt_targets}
\end{align}
where the aggregate interference is computed according to equation \eqref{eq:predicted_interference_refined}.

The resulting input--output pairs provide geometry-consistent, mobility-aware, and regime-aware samples for learning the spatio-temporal evolution of the hybrid-field environment. Physical-network observations periodically recalibrate the DT, while the synchronized context supports fast online GenAI prediction, as illustrated in Fig.~\ref{BD-2}.

\subsection{DT Construction, Calibration, and Online Synchronization}
\label{subsec:dt_construction}

To clarify how the environment information used by the DT is obtained, we distinguish between static environment acquisition, dynamic-state synchronization, and radio calibration. The site-specific DT is initially constructed through a one-time offline acquisition stage, where static information, including room geometry, walls, fixed objects, material properties, and AP locations, can be obtained from available floor plans/computer-aided designs (CAD) or building information modeling (BIM) models, site surveys, or 3D sensing, such as, camera/depth or light detection and ranging (LiDAR)-based scans, and subsequently represented by the geometry and material assignments of the ray-tracing scene. Once constructed, the ray tracer resolves the LoS and reflected paths, from which static blockage conditions and scatterer locations are obtained. Dynamic information, including UE locations and moving blockers or scatterers, is updated through the fast synchronization loop in Fig.~\ref{BD-2} using available localization/tracking information, such as, camera/depth sensing or wireless positioning, together with physical-network measurements. These observations update the dynamic DT state and corresponding blockage and propagation paths without reconstructing the complete static environment at every prediction interval. Finally, measured channel and link-quality information can be used to calibrate discrepancies between the ray-traced DT and the physical network, with persistent environmental changes triggering the slower recalibration path in Fig.~\ref{BD-2}. In our Sionna-based evaluation, the static geometry and material information are directly instantiated by the considered ray-tracing scenes, while mobility and environmental states are varied over time according to the simulated scenario.

Given the DT geometry, the blockage factor in equation \eqref{eq:large_scale_fading_refined} is determined through the geometric relationship between the transmitter, UE, and environmental obstacles as

\begin{equation}
\xi_k(t)=
\begin{cases}
1, & \overline{\mathbf{p}_k\mathbf{u}(t)}
      \text{ is unobstructed},\\
\eta, & \overline{\mathbf{p}_k\mathbf{u}(t)}
      \text{ intersects a blocking object},
\end{cases}
\label{eq:blockage_geometric_rule}
\end{equation}
where the line segment $\overline{\mathbf{p}_k\mathbf{u}(t)}$ represents the direct transmitter--UE path. The intersection test is performed using the same site-specific geometry employed by the ray-tracing engine. Similarly, the scatterer locations $\tilde{\mathbf{p}}_{k,l}$ used in the NLoS channel model are obtained from the reflection points of the resolved ray-tracing paths. Hence, blockage and scatterer information are derived from the DT geometry rather than introduced as independent environmental assumptions. For future states, the GenAI-predicted UE trajectory is evaluated relative to the same geometry, allowing the framework to anticipate changes in the blockage state.

As illustrated by the feedback and recalibration paths in Fig.~\ref{BD-2}, DT synchronization is performed at two timescales. Fast-timescale measurements of channel, UE position, interference, and link quality update the current network state used for GenAI inference, whereas changes in the physical environment that produce persistent disagreement between measured and DT-predicted states trigger slower-timescale DT recalibration. Such recalibration may involve updating the affected geometry/material representation and subsequently refining the predictive model using the updated DT states. This event-driven separation avoids requiring complete high-fidelity ray tracing at every $\Delta t=1$ ms beamforming update while allowing the DT to remain synchronized with slower environmental changes.

\subsection{GenAI-DT Spatio-Temporal Scenario Synthesis}

In addition to baseline data acquisition, the DT--GenAI engine synthesizes diverse spatio-temporal scenarios by integrating physical mobility constraints with environmental context. 
The objective is to expose the predictive model to variations in user mobility, blockage, and user density that jointly drive the non-stationary interference process. This enables both typical operating conditions and less frequent but critical events, such as abrupt mobility changes, blockage transitions, and localized interference surges, to be represented in the training data.

\subsubsection{Predictive Mobility Modeling}

The UE trajectory within the DT evolves according to a state-space representation as
\begin{equation}
{\bf u}(t+1) = {\bf u}(t) + {\bf v}(t)\Delta t + {\boldsymbol \epsilon}(t),
\label{eq:mobility_model}
\end{equation}
where ${\bf v}(t)$ is the velocity vector, $\Delta t$ is the time step, and ${\boldsymbol \epsilon}(t)$ represents small-scale random perturbations. 
Rather than allowing unconstrained mobility evolution, the DT restricts feasible trajectories according to the site-specific geometry $G_{\rm DT}$, thereby preventing predicted motion through physical boundaries and preserving consistency with the indoor layout.
The GenAI module learns this time-correlated process 
from the historical mobility and DT context
to synthesize the future sequence
$\hat{\mathcal{U}}_{\mathrm{future}}=\{{\bf u}(t+\tau)\}_{\tau=1}^{T}$,
providing the spatial context required to predict future hybrid-field channel realizations $\hat{\mathbf{h}}_k(t+\tau)$.
The resulting mobility evolution also determines the future transmitter--UE distances and hence potential NF/FF regime transitions according to equation \eqref{eq:regime_classification}.

\subsubsection{Generative Blockage and Shadowing Dynamics}
\label{sec:blockage}

To model volatile propagation conditions, the DT integrates the predictive blockage state $\hat{\xi}_k(t+\tau)$ into the large-scale fading model as described in equation \eqref{eq:large_scale_fading_refined}.
The geometry-dependent blockage relationship is determined by the DT using the transmitter--UE path and the site-specific obstacles, as described in equation \eqref{eq:blockage_geometric_rule}. Conditioned on this physical relationship and the predicted mobility evolution, the GenAI module learns the temporal transition of blockage states and estimates the future blockage probability in equation \eqref{eq:blockage_prob}.
Consequently, future LoS blockage can be anticipated before the predicted UE trajectory reaches an obstructed region, allowing the corresponding attenuation to be incorporated into the effective channel used for proactive interference management.

\subsubsection{Interference Hotspot Modeling}

To emulate dense user clusters, the DT generates spatial hotspots using a Poisson-based spatial model, which is widely adopted in stochastic-geometry analysis of wireless networks~\cite{andrews_tractable_2011}. Specifically, let $\mathcal{U}_c(t)\subseteq\mathcal{U}$ denote the $c$-th hotspot cluster region at time $t$. The number of users within this region is modeled as
\begin{equation}
|\mathcal{U}_c(t)|
\sim
\mathcal{P}(\lambda_c),
\label{eq:hotspot_model}
\end{equation}
where $\mathcal{P}(\lambda_c)$ denotes a Poisson distribution with mean $\lambda_c$. 
This model is consistent with Poisson-cluster representations of spatial user aggregation in dense cellular networks~\cite{afshang_poisson_2018}. By varying $\lambda_c$ and the hotspot locations over time, the DT generates localized changes in user density that produce corresponding interference surges.
These hotspot realizations complement the mobility and blockage processes by exposing the GenAI model to abrupt changes in aggregate interference caused by spatial user concentration.

\subsubsection{Integrated Spatio-Temporal Data Repository}

By combining the mobility, blockage, and hotspot synthesis processes, the DT constructs a comprehensive spatio-temporal data repository as
\begin{equation}
\widetilde{\mathcal{D}}_{\mathrm{DT}}
=
\left\{
\mathbf{h}_{k,\mathrm{DT}}(t),
I_{\mathrm{DT}}(t),
G_{\mathrm{DT}}(t)
\right\}_{t=1}^{T_{\mathrm{sim}}},
\label{eq:dt_full_dataset}
\end{equation}
where $\mathbf{h}_{k,\mathrm{DT}}(t)$ and $I_{\mathrm{DT}}(t)$ denote the DT-generated channel and corresponding interference state, respectively.
The repository captures the interaction between physical geometry and non-stationary wireless dynamics.
Compared with spatially sparse physical measurements, the DT can systematically generate samples containing less frequent but important conditions, including corner-induced blockage, NF/FF transitions, and dense interference hotspots.
Consequently, it provides a geometry-consistent training substrate for the GenAI module to learn the 
conditional spatio-temporal distribution of future hybrid-field network states used for proactive optimization.

\subsection{DT-Assisted Regime-Aware Channel Modeling}

To accurately synthesize hybrid-field propagation, the DT employs the regime indicator $\rho_k(t)\in\{0,1\}$ to select the appropriate 
distance representation for the FF and NF propagation regimes.
For the LoS component, the distance between the $m$-th antenna element of transmitter $k$ and the UE can be unified as
\begin{equation}
d_{k,0}[m](t)
=
(1-\rho_k(t))\,d_k({\bf u}(t))
+
\rho_k(t)\,\|{\bf u}(t)-{\bf p}_{k,m}\|_2,
\label{eq:los_distance_unified}
\end{equation}
where $d_k({\bf u}(t))$ is the FF distance to the array center and $\|{\bf u}(t)-{\bf p}_{k,m}\|_2$ is the exact NF Euclidean distance to each antenna element.
Thus, $\rho_k(t)=0$ recovers the FF array-center distance approximation, whereas $\rho_k(t)=1$ retains the element-dependent distances required to capture spherical-wave NF propagation.

Similarly, for NLoS components ($l\geq1$), we adopt a single-bounce geometry-based scattering model, commonly used in geometry-based and ray-tracing channel models, where the path length is determined by the transmitter--scatterer and scatterer--UE distances~\cite{jiang_geometry-based_2020, lu_near-field_2023, han_channel_2020}. The propagation distance can be synthesized as
\begin{align}
&d_{k,l}[m](t)
=\\[-1ex] \notag
&(1-\rho_k(t))\,d_{k,l}^{\mathrm{FF}}(t)
+
\rho_k(t)
\left(
\|\mathbf{p}_{k,m}-\tilde{\mathbf{p}}_{k,l}\|_2
+
\|\tilde{\mathbf{p}}_{k,l}-\mathbf{u}(t)\|_2
\right),
\label{eq:nlos_distance_unified}
\end{align}
where $\tilde{\mathbf{p}}_{k,l}$ denotes the coordinates of the $l$-th scatterer provided by the DT geometry.
As discussed in Section~\ref{subsec:dt_construction}, these scatterer locations are obtained from the reflection points of the resolved ray-tracing paths rather than being independently assigned.

These distance formulations are subsequently used to compute the path-wise gains $\beta^{\mathrm{(PD)}}_{k,l}$ under the product-distance path-loss model, ensuring consistency with the NF propagation characteristics described in Section~\ref{Sec:II}.

The regime indicator switches the channel representation between element-dependent spherical-wave NF modeling and array-center FF modeling. As the UE crosses the Rayleigh boundary, the DT updates $\rho_k(t)$ accordingly, enabling geometry-consistent regime-aware prediction and beamforming.

\subsection{GenAI-Driven Spatio-Temporal Prediction}

Using the DT-generated repository $\widetilde{\mathcal{D}}_{\mathrm{DT}}$ in equation \eqref{eq:dt_full_dataset}, the GenAI module is implemented as a cGAN network. 

The cGAN is selected to generate multiple plausible futures under uncertain mobility and blockage while conditioning on site-specific DT, historical, and NF/FF information. Its efficient stochastic sampling also supports the multiple future realizations required by the proactive optimizer. It is adopted for its favorable tradeoff between stochastic conditional generation and online inference complexity. Unlike iterative diffusion sampling, a trained cGAN generates each future realization through a single generator forward pass, enabling efficient sampling of multiple trajectories. Conditional VAEs (CVAEs) provide another efficient alternative but exhibit different diversity--fidelity characteristics. Table~\ref{tab:genai_comparison} summarizes these qualitative tradeoffs.

\begin{table}[t]
\centering
\caption{Qualitative Comparison of Generative Models for the Proposed Prediction Module}
\label{tab:genai_comparison}
\scriptsize
\renewcommand{\arraystretch}{1.05}
\setlength{\tabcolsep}{2.5pt}
\begin{tabular}{|l|c|c|c|c|}
\hline
\textbf{Model} & \textbf{Stochastic} & \textbf{Conditional} &
\textbf{Sampling} & \textbf{Suitability} \\
& \textbf{futures} & \textbf{generation} &
\textbf{cost} & \textbf{here} \\
\hline
cGAN~\cite{goodfellow_generative_2014}
& Yes & Yes & Low & Selected \\
\hline
CVAE~\cite{kingma_auto-encoding_2013}
& Yes & Yes & Low & Alternative \\
\hline
Diffusion~\cite{ho_denoising_2020}
& Yes & Yes & Higher$^\dagger$ & Alternative \\
\hline
Deterministic
& No & Yes & Low & No uncertainty \\
\hline
\multicolumn{5}{l}{\scriptsize
$^\dagger$Conventional iterative diffusion sampling.}
\end{tabular}
\end{table}

The cGAN learns the conditional distribution of future DT network states given the current DT context. Conceptually, the target distribution is

\begin{align}
p_{\mathrm{DT}}\!\left(
\left\{
\mathbf{h}_{k,\mathrm{DT}}(t+\tau)
\right\}_{k\in\mathcal{K}},
I_{\mathrm{DT}}(t+\tau)
\,\middle|\,
{\bf X}^{\mathrm{DT}}(t)
\right),
\end{align}
where $\tau=1,\ldots,T$ is the prediction horizon and ${\bf X}^{\mathrm{DT}}(t)$ denotes the conditioning information extracted from the DT. equation \eqref{eq:dt_features}, the conditioning vector is defined as
\begin{equation}
{\bf c}(t)
=
\Bigl[
{\bf H}_{t-T_h:t}^{\mathrm{hist}},
{\bf u}(t),
\{\rho_k(t)\}_{k\in\mathcal{K}},
\{\xi_k(t)\}_{k\in\mathcal{K}}
\Bigr],
\label{eq:conditioning_vector}
\end{equation}

where the current blockage states $\{\xi_k(t)\}$ are obtained from the synchronized DT geometry as described in Section~\ref{subsec:dt_construction}.

Given the current network context ${\bf c}(t)$, the generator $G(\cdot)$ produces future channel and interference trajectories. To capture uncertainty in future evolution, a random noise vector ${\bf z}\sim p_{\bf z}$ is introduced, 
where $p_{\bf z}$ is chosen as a Gaussian latent distribution. Different realizations of ${\bf z}$ under the same conditioning context enable the generator to represent multiple plausible future outcomes.
The generated trajectory is expressed as
\begin{align}
\hat{\mathcal{Y}}(t) =
\left[
\left\{
\hat{\bf h}_k(t+\tau;{\bf z},{\bf c}(t))
\right\}_{k\in\mathcal{K}},
\hat{I}(t+\tau;{\bf z},{\bf c}(t))
\right]_{\tau=1}^{T},
\label{eq:generator_output}
\end{align}
where the predicted channels $\hat{\bf h}_k(t+\tau)$ and interference $\hat{I}(t+\tau)$ are conditioned on ${\bf z}$ and ${\bf c}(t)$ through the generator.
Repeated sampling of ${\bf z}$ produces the $M$ future realizations subsequently used by the proactive optimizer.

To ensure that the generated trajectories follow the DT-derived distribution, the discriminator $D(\cdot)$ distinguishes between DT-generated target trajectories and synthetic trajectories produced by $G(\cdot)$ under the same conditioning context. 
Specifically, $D(\mathcal{Y}\mid{\bf c}(t))\in[0,1]$ represents the discriminator output associated with trajectory $\mathcal{Y}$ conditioned on ${\bf c}(t)$.
The adversarial training objective is
\begin{align}
\mathcal{L}_{\mathrm{adv}} =  \notag
&\min_G \max_D V(D,G) \\ \notag
&= \mathbb{E}_{\mathcal{Y}\sim p_{\mathrm{DT}}(\cdot \mid {\bf c}(t))} 
\big[ \log D(\mathcal{Y}\mid{\bf c}(t)) \big] \nonumber \\
&\quad + \mathbb{E}_{{\bf z}\sim p_{\bf z}} 
\big[ \log\big(1-D(G({\bf z},{\bf c}(t))\mid{\bf c}(t))\big) \big],
\label{eq:gan_loss}
\end{align}
where $\mathcal{L}_{\mathrm{adv}}$ denotes the adversarial loss.

Adversarial consistency alone does not guarantee that the generated trajectories preserve the geometry-dependent channel and interference evolution required by the physical model. Therefore, the generator is additionally regularized using the prediction-consistency loss given in equation \eqref{eq:prediction_loss}.

\begin{figure*}[t!]
\begin{align}
\mathcal{L}_{\mathrm{pred}}
=
\sum_{\tau=1}^{T}
\left(
\sum_{k\in\mathcal{K}}
\left\|
\hat{\bf h}_k(t+\tau)
-
{\bf h}_{k,\mathrm{DT}}(t+\tau)
\right\|_2^2
+
\mu
\left|
\hat{I}(t+\tau)
-
I_{\mathrm{DT}}(t+\tau)
\right|^2
\right),
\label{eq:prediction_loss}
\end{align}
\normalsize
\hrulefill
\end{figure*}

$\mu>0$ in equation \eqref{eq:prediction_loss} controls the relative contribution of interference-prediction fidelity. The use of DT-generated targets constrains the learned future states toward the geometry-consistent propagation behavior established in Section~\ref{Sec:II}.

The overall training objective is
\begin{equation}
\mathcal{L}_{\mathrm{GenAI}}
=
\mathcal{L}_{\mathrm{adv}}
+
\lambda_{\mathrm{pred}}\mathcal{L}_{\mathrm{pred}},
\label{eq:overall_genai_loss}
\end{equation}
where $\lambda_{\mathrm{pred}}$ is a regularization weight.

The adversarial term promotes distributional realism, whereas the prediction term promotes consistency with the DT-generated channel and interference trajectories. After training, stochastic latent sampling enables the generator to produce multiple future realizations for robust proactive optimization. The diversity of these generated trajectories is quantitatively evaluated in Section~\ref{Sec:V} to assess whether the model captures distinct mobility and blockage outcomes rather than collapsing to a single future mode.
Algorithm~\ref{alg:genai_predict} summarizes the stochastic future-state generation process.

\begin{algorithm}[!t]
\caption{GenAI-Based Future Trajectory}
\label{alg:genai_predict}
\begin{algorithmic}[1]
\STATE \textbf{Input:} ${\bf X}^{\mathrm{DT}}(t)$, $G(\cdot)$, $M$, $T$
\STATE \textbf{Output:} $\bigl\{\hat{\mathcal{Y}}^{(m)}(t)\bigr\}_{m=1}^{M}$

\STATE Extract ${\bf c}(t)$ from ${\bf X}^{\mathrm{DT}}(t)$ using \eqref{eq:conditioning_vector}

\FOR{$m=1$ to $M$}
    \STATE Sample ${\bf z}^{(m)} \sim p_{\bf z}$
    \STATE Generate future states using \eqref{eq:generator_output}
    \STATE Obtain $\left\{\hat{\mathbf{h}}_k^{(m)}(t+\tau), \hat{\Lambda}_k^{(m)}(t+\tau), \hat{\xi}_k^{(m)}(t+\tau)\right\}_{k\in\mathcal{K},\,\tau=1}^{T}$ using \eqref{eq:genai_mapping_refined}
    \STATE Form $\left\{\hat{\mathbf{h}}_k^{(m),\mathrm{eff}}(t+\tau)\right\}_{k\in\mathcal{K},\,\tau=1}^{T}$ using \eqref{eq:effective_channel_refined}
    \STATE Compute  \eqref{eq:predicted_interference}
    \STATE Compute \eqref{eq:proactive_sinr_refined}
\ENDFOR

\STATE Return $\bigl\{\hat{\mathcal{Y}}^{(m)}(t)\bigr\}_{m=1}^{M}$ to the proactive optimizer
\end{algorithmic}
\end{algorithm}

\subsubsection{Diversity and Event Coverage of Generated Trajectories}
\label{subsec:diversity}

Generating multiple stochastic realizations is useful for robust optimization only if the latent samples produce distinct plausible futures rather than near-identical trajectories. To quantitatively assess this property, we evaluate the diversity of the $M$ generated mobility trajectories using the average pairwise displacement, which can be written as

\begin{equation}
D_{\mathrm{div}}
=
\frac{2}{M(M-1)T}
\sum_{i<j}
\sum_{\tau=1}^{T}
\left\|
\hat{\mathbf{u}}^{(i)}(t+\tau)
-
\hat{\mathbf{u}}^{(j)}(t+\tau)
\right\|_2 ,
\label{eq:trajectory_diversity}
\end{equation}

where $\hat{\mathbf{u}}^{(i)}(t+\tau)$ denotes the predicted UE position at horizon $\tau$ for the $i$-th generated realization. A value close to zero indicates limited sensitivity to the latent input and potentially insufficient trajectory diversity, whereas larger values indicate greater dispersion among plausible future mobility realizations. The resulting diversity is evaluated jointly with prediction accuracy in Section~\ref{Sec:V}, since diversity alone is not sufficient if the generated trajectories are physically implausible or inaccurate.

In addition to spatial diversity, we evaluate whether the generated realizations capture propagation events that directly affect proactive interference management. For each conditioning context, the generated set is therefore examined for the occurrence of alternative blockage states and NF/FF regimes over the prediction horizon. We define the event-coverage ratio as

\begin{equation}
C_{\mathrm{event}}
=
\frac{N_{\mathrm{covered}}}{N_{\mathrm{ref}}},
\qquad
0\le C_{\mathrm{event}}\le1,
\label{eq:event_coverage}
\end{equation}

where $N_{\mathrm{ref}}$ is the number of distinct relevant future event states represented in the corresponding DT reference samples and $N_{\mathrm{covered}}$ is the number of those states reproduced by at least one of the $M$ generated trajectories. Relevant events include blockage/no-blockage outcomes and NF/FF regime transitions. Section~\ref{Sec:V} reports both $D_{\mathrm{div}}$ and $C_{\mathrm{event}}$ and illustrates representative generated trajectories around blockage and regime-transition regions.

\subsection{Proactive Interference-Aware Optimization}
\label{sec:proactive_opt}

The objective of the proposed framework is to move beyond reactive interference mitigation by leveraging the synthesized spatio-temporal dynamics provided by the GenAI--DT engine. 
Rather than optimizing for a single instantaneous or deterministic future channel, the proposed design accounts for multiple plausible future network states when determining the beamforming vectors.

The GenAI module generates a set of $M$ 
stochastic future realizations
by sampling the latent variable ${\bf z}$, where $M$ denotes the number of sampled future trajectories. The $m$-th synthesized channel trajectory is defined as
\begin{equation}
\hat{\mathcal{H}}^{(m)}
=
\left\{
\hat{\bf h}_k^{(m)}(t+\tau)
\right\}_{k\in\mathcal{K},\,\tau=1}^{T},
\label{eq:predicted_channel_trajectory}
\end{equation}
where $\hat{\mathcal{H}}^{(m)}$ corresponds to the channel component of the $m$-th synthesized future trajectory, and $m\in\{1,\ldots,M\}$ indexes the possible future states.
The diversity and event coverage of these realizations are evaluated in Section~\ref{subsec:diversity}, ensuring that the optimizer is not supplied with near-identical replicas of a single forecast.

Based on these realizations, the predicted aggregate interference at time $t+\tau$ for a given beamforming policy $\{{\bf w}_k\}$ is
\begin{equation}
\hat{I}^{(m)}(t+\tau)
=
\sum_{k\in\mathcal{K}_I}
P_k
\left|
\left(
\hat{\mathbf h}_k^{(m),\mathrm{eff}}(t+\tau)
\right)^H
\mathbf w_k
\right|^2,
\label{eq:predicted_interference}
\end{equation}
where $\hat{\mathbf h}_k^{(m),\mathrm{eff}}(t+\tau)$ incorporates the predicted hybrid-field channel structure, path loss, and blockage attenuation for realization $m$.

To account for uncertainty induced by mobility, blockage, and hybrid-field evolution, the proactive beamforming design minimizes the 
sample-average cumulative interference
across the generated futures:
\begin{subequations}
\begin{align}
\min_{\{{\bf w}_k\}} \quad 
& \frac{1}{M}
\sum_{m=1}^{M}
\sum_{\tau=1}^{T}
\hat{I}^{(m)}(t+\tau),
\label{eq:proactive_opt_final} \\
\text{s.t.} \quad 
&
\frac{
P_0
\left|
\left(
\hat{\bf h}_0^{(m),\mathrm{eff}}(t+\tau)
\right)^H
{\bf w}_0
\right|^2
}{
\hat{I}^{(m)}(t+\tau)+\sigma^2
}
\geq \gamma_{\min},
\quad \forall \tau,m,
\label{eq:sinr_constraint_robust}\\
&
\sum_{k\in\mathcal{K}}
P_k\|{\bf w}_k\|^2
\leq P_{\mathrm{total}}.
\label{eq:power_constraint_robust}
\end{align}
\end{subequations}

The objective in equation \eqref{eq:proactive_opt_final} minimizes the empirical expectation of cumulative interference over the $M$ generated future realizations, while equation \eqref{eq:sinr_constraint_robust} enforces the required SINR for each sampled realization and prediction step. The use of $\hat{\bf h}_k^{(m),\mathrm{eff}}(t+\tau)$ maintains consistency with the system model by jointly accounting for hybrid-field propagation, large-scale attenuation, and predicted blockage. Consequently, the beamforming vectors are optimized against a distribution of plausible future conditions rather than a single forecast. Their robustness to less frequent blockage and mobility events therefore depends on the corresponding events being represented among the generated trajectories, which is quantified through the diversity and event-coverage analysis in Section~\ref{subsec:diversity}.

As summarized in Algorithm~\ref{alg:proactive_opt}, evaluating the interference terms over $M$ realizations, $T$ prediction steps, and $|\mathcal{K}|$ links scales as $\mathcal{O}(MT|\mathcal{K}|)$ per optimization iteration, excluding the cost of the beamforming update itself. This scaling characterizes the trajectory-evaluation component and should not be interpreted as the end-to-end complexity of the complete DT--GenAI pipeline. As illustrated in Fig.~\ref{BD-2}, high-fidelity DT construction and ray-tracing calibration operate at the slower timescale, whereas GenAI inference and proactive optimization constitute the fast online loop. Model parameters, FLOPs, and measured online latency are reported in Section~\ref{Sec:V}.

\begin{algorithm}[!t]
\caption{Proactive Interference-Aware Beamforming Optimization}
\label{alg:proactive_opt}
\begin{algorithmic}[1]

\STATE \textbf{Input:} 
$\{\hat{\mathcal{Y}}^{(m)}(t)\}_{m=1}^{M}$,
predicted UE states,
power constraint $P_{\mathrm{total}}$, $\gamma_{\min}$

\STATE \textbf{Output:} 
$\{{\bf w}_k^{\mathrm{pro}}\}_{k\in\mathcal{K}}$

\STATE Initialize $\{{\bf w}_k^{(0)}\}$ using the predicted NF/FF propagation conditions

\FOR{each link $k\in\mathcal{K}$}

    \STATE Evaluate the Rayleigh-distance criterion in equation
    \eqref{eq:regime_classification} at the predicted UE position

    \IF{$d_k({\bf u}(t+\tau))>R_k$}
        \STATE Initialize ${\bf w}_k^{(0)}$ using FF ZF beamforming~\cite{spencer_zero-forcing_2004}
    \ELSE
        \STATE Initialize ${\bf w}_k^{(0)}$ using NF location-aware beam focusing~\cite{li_multi-user_2024}
    \ENDIF

\ENDFOR

\REPEAT

    \STATE Evaluate $\hat{I}^{(m)}(t+\tau)$ using equation
    \eqref{eq:predicted_interference} for all $m$ and $\tau$

    \STATE Update $\{{\bf w}_k\}$ to minimize the
    sample-average interference in equation \eqref{eq:proactive_opt_final}

    \STATE Enforce the SINR constraints in equation
    \eqref{eq:sinr_constraint_robust} and the total-power constraint in equation \eqref{eq:power_constraint_robust}

\UNTIL{convergence or maximum number of iterations is reached}

\STATE Set
$\{{\bf w}_k^{\mathrm{pro}}\}
\leftarrow
\{{\bf w}_k\}$

\STATE Deploy $\{{\bf w}_k^{\mathrm{pro}}\}$ in the physical network

\STATE Acquire physical-network feedback and synchronize the DT--GenAI state

\end{algorithmic}
\end{algorithm}

\subsubsection{Multi-Timescale Operation, Complexity, and Latency}
\label{subsec:complexity_latency}

The proposed framework adopts the multi-timescale operation illustrated in Fig.~\ref{BD-2} to avoid executing computationally intensive high-fidelity ray tracing at every beamforming update. At the slow timescale, the DT performs site-specific construction, ray-tracing calibration, environment updates, and training-data generation, whereas the fast timescale uses the synchronized DT context for cGAN inference, effective-channel construction, and proactive beamforming optimization. Thus, complete ray tracing is not required within the online control loop.

Table~\ref{tab:complexity_latency} summarizes the computational complexity and measured latency of these operations. The GenAI complexity is characterized by the trainable parameters and FLOPs of the cGAN, with online inference primarily determined by the generator since the discriminator is used only during training. The FLOP count is analytically estimated for the fully connected layers using two floating-point operations per multiply--accumulate (MAC). The generator cost increases approximately linearly with the number of sampled futures $M$, introducing a performance--complexity tradeoff between broader future-state coverage and online computation. The sensitivity to $M$ and the resulting communication-performance/runtime tradeoff are evaluated in Section~\ref{Sec:V}.

\begin{table}[t]
\centering
\caption{Computational Complexity and Latency of the Proposed Framework}
\label{tab:complexity_latency}
\scriptsize
\renewcommand{\arraystretch}{1.03}
\setlength{\tabcolsep}{2.5pt}
\begin{tabular}{|l|c|c|}
\hline
\textbf{Component / Metric} & \textbf{Timescale} & \textbf{Value} \\
\hline
DT construction / ray tracing & Slow & $8.80$ ms \\
\hline
DT synchronization & Fast & $0.04$ ms \\
\hline
cGAN inference ($M=10$) & Fast & $0.63$ ms \\
\hline
Effective-channel construction & Fast & $0.63$ ms \\
\hline
Beamforming optimization & Fast & $14.27$ ms \\
\hline
\textbf{Total online control} & \textbf{Fast} & $\mathbf{15.57}$ ms \\
\hline
Generator architecture & Training/Inference & $[256,512,512]$ \\
\hline
Discriminator architecture & Training & $[512,256,128]$ \\
\hline
Generator parameters & Training/Inference & $428{,}830$ \\
\hline
Discriminator parameters & Training & $204{,}801$ \\
\hline
Generator FLOPs / inference & Fast & $\approx 8.50\times10^{5}$ \\
\hline
Online GenAI FLOPs ($M=10$) & Fast & $8{,}499{,}200$ \\
\hline
Optimizer iterations & Fast & $30$ \\
\hline
Hardware platform & -- & Cloud GPU, 16 GB RAM \\
\hline
\end{tabular}
\end{table}

\subsection{Regime-Aware Precoding Strategy}

The proactive optimization in equation \eqref{eq:proactive_opt_final} exploits the distinct spatial characteristics of FF and NF propagation rather than applying a common precoder to all links. For each predicted future state, the propagation condition is determined by applying the Rayleigh-distance criterion in equation \eqref{eq:regime_classification} to the predicted UE location. The corresponding precoding strategy is selected as follows:

\begin{itemize}

\item \textbf{Far-field regime:}
Conventional angular-domain beamforming, 
represented here by ZF precoding,
is employed based on the predicted FF channel matrix~\cite{spencer_zero-forcing_2004}.

\item \textbf{Near-field regime:}
Location-aware beamforming is adopted using the spherical-wave channel model, enabling spatial energy focusing at the predicted UE location and thereby exploiting both angular and distance-domain information~\cite{li_multi-user_2024}.

\item \textbf{Hybrid regime:}

When NF and FF links coexist, the links are partitioned according to their predicted propagation conditions and jointly optimized using their corresponding effective channels. This allows the optimization in equation \eqref{eq:proactive_opt_final} to account for inter-regime interference while satisfying the SINR and total-power constraints.

\end{itemize}

Thus, regime awareness affects not only the classification of each link but also the underlying channel representation and precoder initialization: FF links employ angular-domain processing, whereas NF links retain the element-dependent spherical-wave structure required for location-domain focusing. For generated futures near the Rayleigh boundary, different realizations may yield different NF/FF outcomes; the robust optimization retains all $M$ effective-channel realizations, thereby accounting for uncertainty associated with the predicted regime transition. The resulting vectors $\{\mathbf w_k^{\mathrm{pro}}\}$ are obtained from equations \eqref{eq:proactive_opt_final}--\eqref{eq:sinr_constraint_robust}, thereby accounting for uncertainty across the $M$ predicted future realizations.

\section{Simulation Setup and Results}
\label{Sec:V}

This section evaluates the proposed GenAI-enhanced DT framework for proactive interference management in dense indoor XL-MIMO systems. 
We first describe the ray-tracing environments and verify that the considered deployment contains both NF and FF links. We then evaluate the proposed framework under different mobility, blockage, user-density, and prediction conditions and compare it with representative reactive and predictive baselines.

\subsection{Simulation Scenario}

\begin{table}[!t]
\centering
\caption{Simulation Parameters}
\label{tab:sim_params}
\scriptsize
\renewcommand{\arraystretch}{1.05}
\setlength{\tabcolsep}{3.5pt}
\begin{tabular}{|l|c|}
\hline
\textbf{Parameter} & \textbf{Value} \\
\hline
Carrier frequency $f_c$ & $100$ GHz (D-band) \\
\hline
Bandwidth & $400$ MHz \\
\hline
UPA size ($N_x\times N_y$) & $16\times16=256$ \\
\hline
Element spacing $d$ &
$5$ mm ($\approx1.67\lambda$) \\
\hline
Array aperture $D_k$ (diagonal) &
$\approx0.106$ m \\
\hline
Rayleigh distance $R_k$ &
$\approx7.5$ m \\
\hline
Number of users $K$ & $2$--$12$~\cite{alkhateeb_limited_2015} \\
\hline
Coherence time $T_c$ & $\approx1.27$ ms \\
\hline
Prediction step $\Delta t$ & $1$ ms \\
\hline
Prediction horizon $T$ & $5$ steps ($5$ ms) \\
\hline
Generated trajectories $M$ & $10$ \\
\hline
Maximum UE speed $v_{\max}$ & $1$ m/s \\
\hline
Noise power $\sigma^2$ & $-80$ dBm \\
\hline
Transmit power $P_k$ & $0$ dBm (normalized) \\
\hline
SINR threshold $\gamma_{\min}$ & $5$ dB \\
\hline
Blockage factor $\eta$ & $0.01$~\cite{rappaport_millimeter_2013} \\
\hline
Hotspot intensity $\lambda_c$ & $3$ users/cluster~\cite{haenggi_stochastic_2012} \\
\hline
Simulation duration $T_{\mathrm{sim}}$ & $100$ ms \\
\hline
\end{tabular}
\end{table}

The main simulation parameters are summarized in Table~\ref{tab:sim_params}. 
At $f_c=100$ GHz, the carrier wavelength is $\lambda\approx3$ mm. For the considered $16\times16$ UPA with element spacing $d=5$ mm, the maximum diagonal aperture is approximately
$D_k=\sqrt{2}(N_x-1)d\approx0.106$ m. Accordingly, equation \eqref{eq:rayleigh_distance} gives a Rayleigh distance of $R_k\approx7.5$ m. The enlarged spacing is adopted here to realize a sparse large-aperture XL-MIMO configuration and thereby obtain an indoor-scale NF region with a fixed number of antenna elements.

\subsubsection{Multi-Scene Test Environments}
\label{subsec:multiscene}

\begin{figure*}[t]
     \centering
     \begin{subfigure}[b]{0.37\textwidth}
         \centering
         \includegraphics[width=\textwidth]{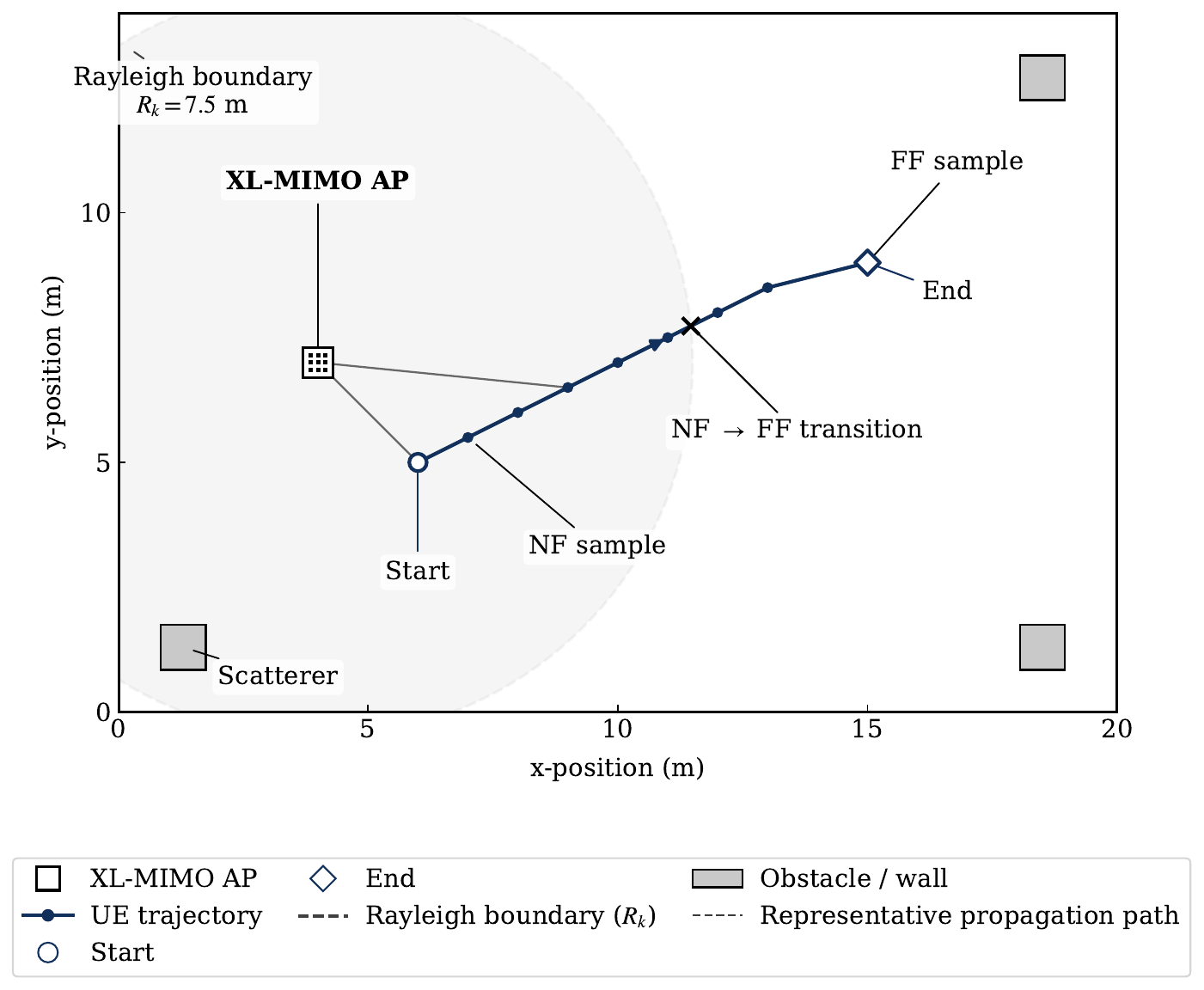}
         \vspace{-1.5ex}
         \caption{Scene A: Open room with predominantly LoS propagation.}
         \label{test_scene_a}
     \end{subfigure}
     \hfill
     \begin{subfigure}[b]{0.37\textwidth}
         \centering
         \includegraphics[width=\textwidth]{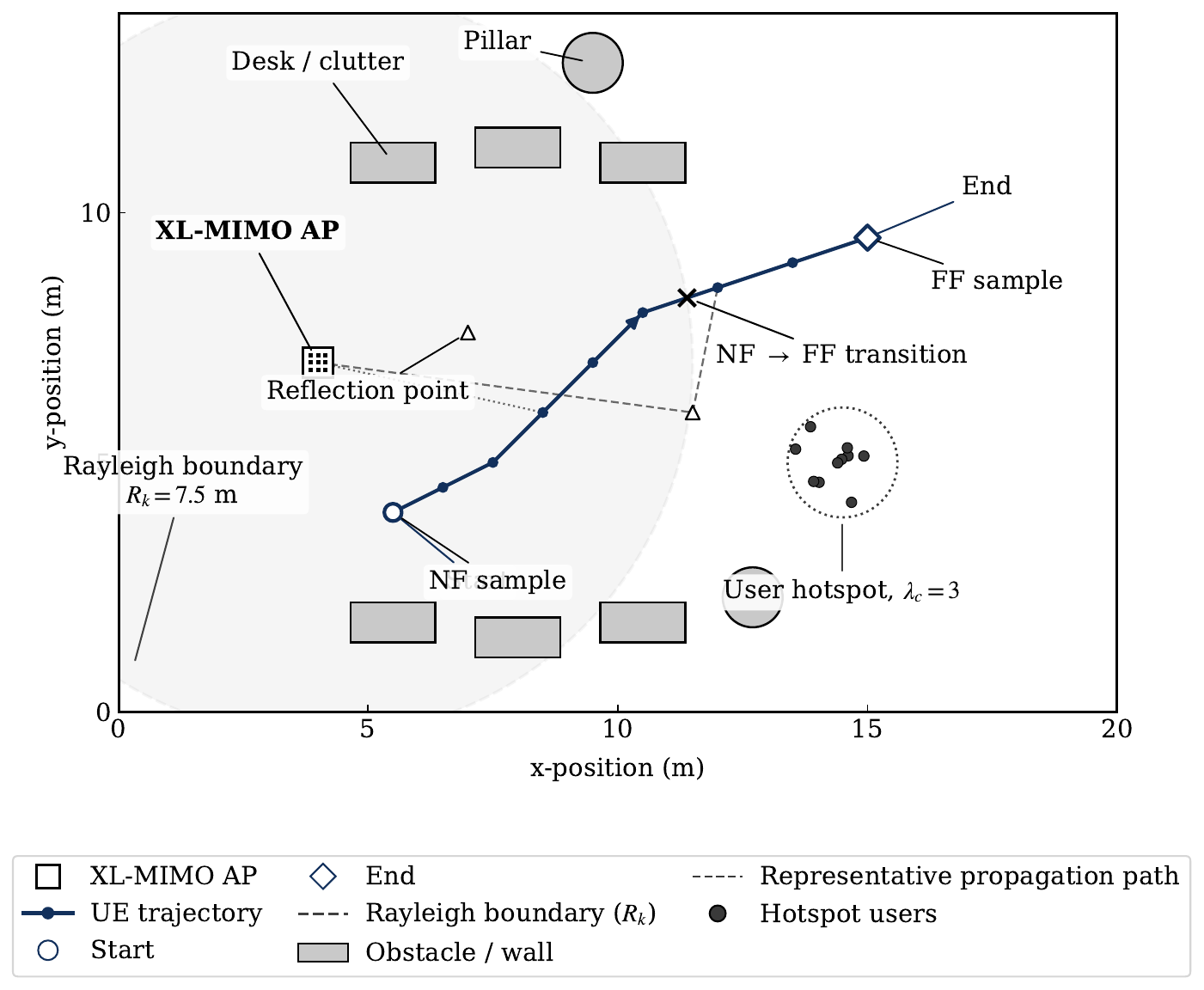}
         \vspace{-1.5ex}
         \caption{Scene B: Dense cluttered environment with a user hotspot.}
         \label{test_scene_b}
     \end{subfigure}
     \caption{Sionna ray-tracing environments used for multi-scene evaluation.}
     \vspace{-1.5ex}
     \label{fig:test_scenes}
\end{figure*}

To evaluate the generalization of the proposed framework across different propagation environments, we consider two geometrically distinct indoor ray-tracing scenes, as illustrated in Fig.~\ref{fig:test_scenes}. Fig.~\ref{test_scene_a} depicts scene A with an open indoor environment with predominantly LoS propagation and limited obstruction. On the contrary, Fig.~\ref{test_scene_b} depicts scene B, which considers a dense cluttered environment with multiple obstacles and a user hotspot, introducing stronger blockage, scattering, and interference dynamics. These contrasting environments enable evaluation of the proposed framework under both relatively favorable and challenging indoor propagation conditions.

For each scene, Fig.~\ref{fig:test_scenes} shows the AP location, UE trajectory, principal obstacles/scatterers, and the NF/FF boundary corresponding to $R_k$. The trajectories span both sides of the Rayleigh boundary, producing NF, FF, and regime-transition samples. The corresponding transmitter--UE distance ranges and NF/FF sample proportions are summarized in Table~\ref{tab:hybrid_validation}.

\subsubsection{Hybrid-Field Validity of the Simulated Deployment}
\label{subsec:hybrid_validity}

To verify that the considered deployment genuinely contains both NF and FF links, we evaluate the simulated transmitter--UE distances relative to the Rayleigh distance in equation \eqref{eq:rayleigh_distance}. At $f_c=100$ GHz, the carrier wavelength is $\lambda\approx3$ mm. For the considered $16\times16$ UPA with element spacing $d=5$ mm, the physical extent between the outermost elements is $D_x=D_y=(N_x-1)d=75$ mm, yielding a maximum diagonal aperture of $D_k=\sqrt{D_x^2+D_y^2}\approx0.106$ m. Consequently, the corresponding Rayleigh distance is $R_k\approx7.5$ m. Table~\ref{tab:hybrid_validation} reports the resulting distance range and NF/FF sample proportions for each scene.

\begin{table}[t]
\centering
\caption{Hybrid-Field Characteristics of the Test Scenes}
\label{tab:hybrid_validation}
\scriptsize
\renewcommand{\arraystretch}{1.05}
\setlength{\tabcolsep}{3pt}
\begin{tabular}{|l|c|c|c|}
\hline
\textbf{Scene} &
\textbf{AP--UE Distance (m)} &
\textbf{NF Samples (\%)} &
\textbf{FF Samples (\%)} \\
\hline
Open room & $0.7$--$17.0$ & $52.3$ & $47.7$ \\
\hline
Dense/cluttered & $0.7$--$17.0$ & $51.8$ & $48.2$ \\
\hline
\end{tabular}
\end{table}

\subsection{Benchmark Schemes}

To evaluate the effectiveness of the proposed GenAI-enhanced DT framework, we compare it against the following benchmark schemes:

\begin{enumerate}

\item \textbf{FF Reactive ZF~\cite{spencer_zero-forcing_2004}:}
A conventional reactive beamforming scheme that computes ZF precoders using only the instantaneous channel state information (CSI) at time $t$. This benchmark assumes FF propagation and does not exploit prediction or DT assistance.

\item \textbf{Reactive Hybrid-Field (Regime-Aware)~\cite{hu_near-field_2022}:}
A regime-aware reactive baseline in which FF links are served using angular-domain ZF precoding, whereas NF links employ location-aware beam focusing. 
This benchmark incorporates NF/FF propagation awareness but relies only on the instantaneous channel state, thereby isolating the gain provided by proactive future-state prediction.

\item \textbf{Deterministic DT Baseline (Non-Generative DT)~\cite{khan_digital_2026}:}
A predictive baseline that uses the DT to generate a single deterministic future trajectory based on the estimated UE motion and environmental evolution. The beamforming design is optimized using this single predicted trajectory without stochastic latent-variable sampling. 
This benchmark represents non-generative DT-assisted prediction and isolates the benefit of generating multiple plausible future realizations under mobility and blockage uncertainty.

\item \textbf{GenAI-Enhanced DT Regime-Unaware~\cite{li_vehicle_2021}:}
A proactive benchmark that uses the cGAN-generated future trajectories but applies a unified beamforming strategy without distinguishing between NF and FF links. 
This benchmark retains stochastic future-state prediction while removing propagation-regime awareness, thereby isolating the contribution of the proposed NF/FF-aware channel modeling and precoding strategy.

\item \textbf{Proposed GenAI-Enhanced DT Framework (Regime-Aware):}
The proposed framework combines site-specific DT-assisted environmental modeling, stochastic multi-trajectory prediction using the cGAN, 
geometry-aware blockage prediction, and NF/FF regime-aware proactive beamforming over multiple plausible future network states.

\end{enumerate}

\subsection{Mobility-Speed Generalization Study}
\label{subsec:mobility}

Although the nominal configuration considers $v_{\max}=1$ m/s, we further evaluate the framework at $v\in\{0.5,1,1.5,2\}$ m/s in the more challenging Scene~B. Interference-prediction RMSE is measured under the same framework configuration to assess sensitivity to increasing UE mobility.

\subsection{Numerical Results and Analysis}

We first evaluate the proposed framework in the open-room environment depicted in Fig.~\ref{test_scene_a} and then examine its performance in the more challenging cluttered/hotspot environment, illustrated in Fig.~\ref{test_scene_b}. The two scenarios assess whether the gains of proactive prediction and NF/FF-aware processing persist under different propagation and interference conditions.

\begin{figure*}[t]
\centering

\begin{subfigure}[b]{0.32\textwidth}
    \centering
    \includegraphics[width=\textwidth]{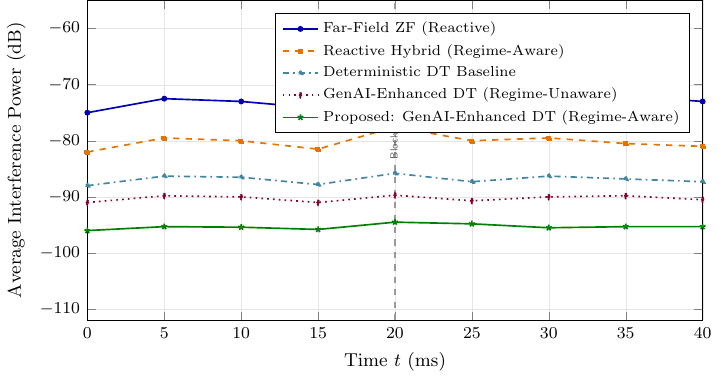}
    \caption{Interference power versus time.}
    \label{fig:sceneA_time}
\end{subfigure}
\hfill
\begin{subfigure}[b]{0.32\textwidth}
    \centering
    \includegraphics[width=\textwidth]{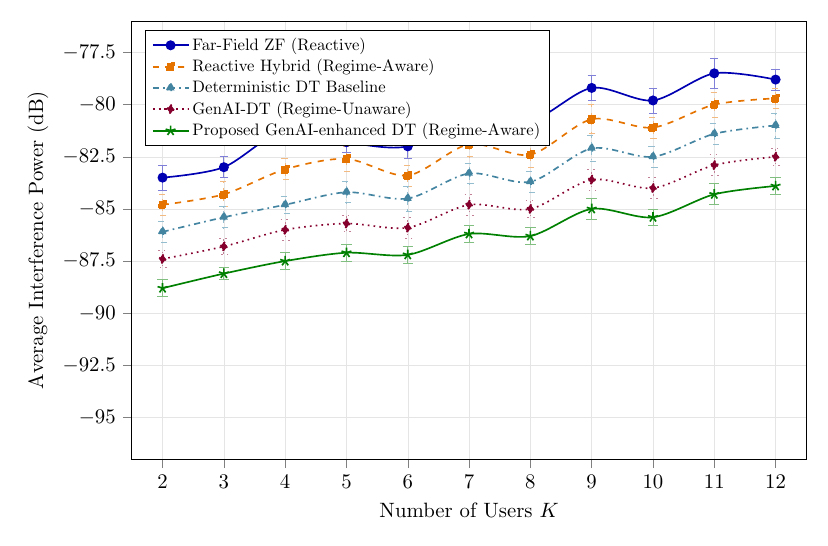}
    \caption{Interference power versus users.}
    \label{fig:sceneA_users}
\end{subfigure}
\hfill
\begin{subfigure}[b]{0.32\textwidth}
    \centering
    \includegraphics[width=\textwidth]{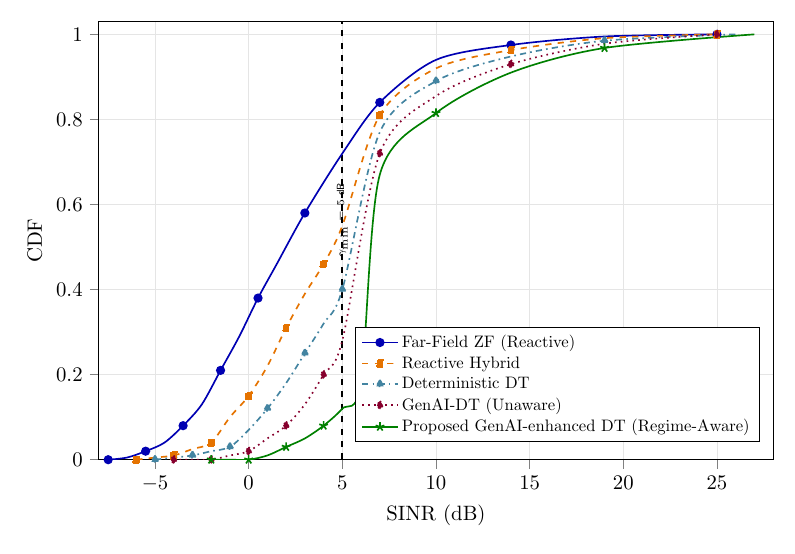}
    \caption{CDF of achievable SINR.}
    \label{fig:sceneA_cdf}
\end{subfigure}

\caption{
Performance comparison in Scene~A: (a) temporal interference,
(b) interference versus number of users, and (c) SINR cumulative distribution function (CDF).
}
\label{fig:sceneA_results}
\end{figure*}

\begin{figure*}[t]
\centering

\begin{subfigure}[b]{0.32\textwidth}
    \centering
    \includegraphics[width=\textwidth]{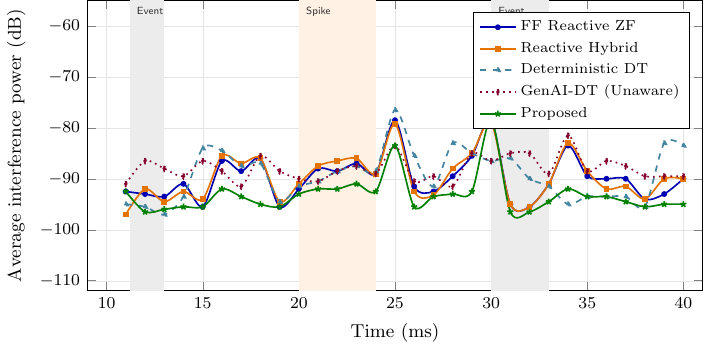}
    \caption{Interference power versus time.}
    \label{fig:sceneB_time}
\end{subfigure}
\hfill
\begin{subfigure}[b]{0.32\textwidth}
    \centering
    \includegraphics[width=\textwidth]{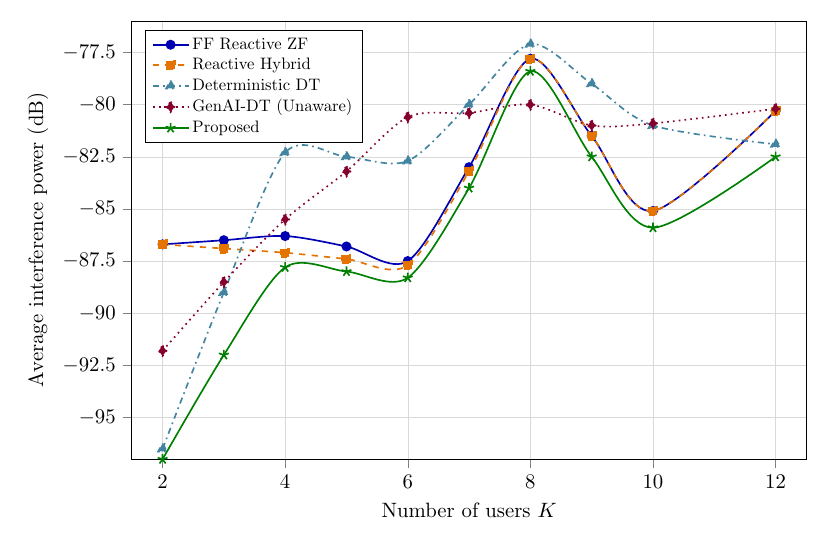}
    \caption{Interference power versus users.}
    \label{fig:sceneB_users}
\end{subfigure}
\hfill
\begin{subfigure}[b]{0.32\textwidth}
    \centering
    \includegraphics[width=\textwidth]{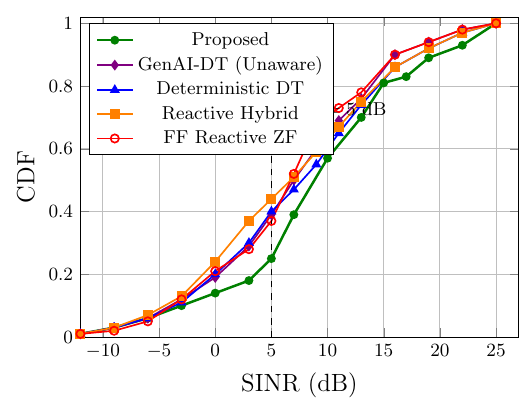}
    \caption{CDF of achievable SINR.}
    \label{fig:sceneB_cdf}
\end{subfigure}

\caption{
Performance comparison in Scene~B: (a) temporal interference,
(b) interference versus number of users, (c) SINR CDF.
}
\label{fig:sceneB_results}
\end{figure*}

Fig.~\ref{fig:sceneA_results} presents the results for Scene~A. In Fig.~\ref{fig:sceneA_time}, the proposed framework maintains the lowest interference over time, including around the indicated blockage event, whereas the reactive schemes experience higher interference. Fig.~\ref{fig:sceneA_users} shows increasing interference with user density for all schemes, while the proposed method consistently provides the lowest values over $K=2$--$12$. The SINR CDF in Fig.~\ref{fig:sceneA_cdf} is also shifted furthest to the right for the proposed method, yielding the smallest probability of operating below $\gamma_{\min}=5$ dB. These results demonstrate the complementary gains of future-state prediction and NF/FF-aware proactive beamforming in the relatively open environment.

Fig.~\ref{fig:sceneB_results} evaluates the framework in the cluttered/hotspot environment, where blockage, reflected propagation, and spatial user concentration produce stronger interference variations. Fig.~\ref{fig:sceneB_time} shows pronounced interference changes around the indicated event intervals. Nevertheless, the proposed scheme generally maintains lower interference than the reactive baselines. Fig.~\ref{fig:sceneB_users} further illustrates the more irregular dependence on user density caused by the scene geometry and hotspot distribution, while the proposed method remains competitive and achieves particularly low interference over most operating points. Finally, Fig.~\ref{fig:sceneB_cdf} shows improved lower-tail SINR for the proposed framework, including a smaller CDF value around the operating threshold $\gamma_{\min}=5$ dB. These results demonstrate that the proposed gains are retained under substantially more complex propagation conditions than in Scene~A.

\begin{figure*}[t]
\centering

\begin{subfigure}[b]{0.33\textwidth}
    \centering
    \includegraphics[width=\textwidth]{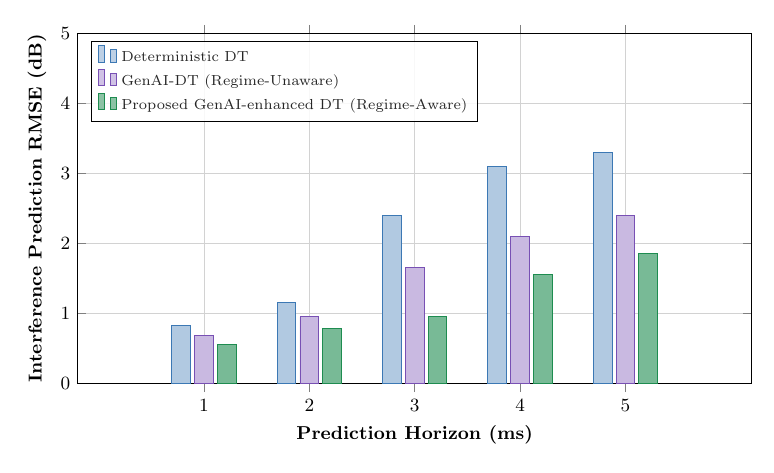}
    \caption{Scene~A.}
    \label{fig:rmse_sceneA}
\end{subfigure}
\hfill
\begin{subfigure}[b]{0.33\textwidth}
    \centering
    \includegraphics[width=\textwidth]{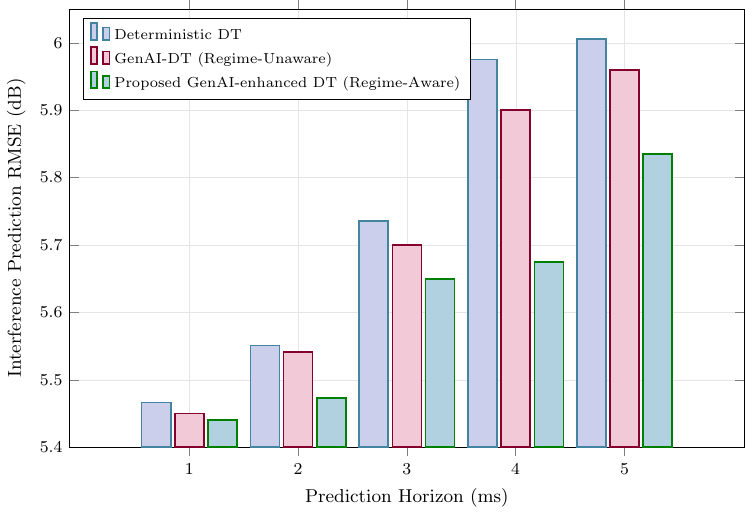}
    \caption{Scene~B.}
    \label{fig:rmse_sceneB}
\end{subfigure}

\caption{
Interference-prediction RMSE versus prediction horizon in
(a) Scene~A and (b) Scene~B.
}
\label{fig:rmse_scenes}
\end{figure*}

Fig.~\ref{fig:rmse_scenes} evaluates prediction accuracy over horizons of $1$--$5$ ms. In both scenes, RMSE increases with the prediction horizon as uncertainty accumulates, while the proposed regime-aware GenAI-DT consistently achieves the lowest error among the predictive schemes. Scene~B exhibits substantially larger RMSE than Scene~A because its clutter, blockage, and hotspot dynamics create a more difficult prediction environment. Nevertheless, the proposed method retains its advantage over both deterministic DT and regime-unaware GenAI-DT prediction, demonstrating the benefit of stochastic generation and hybrid-field awareness across different environments.

\begin{figure}[t]
\centering
\includegraphics[width=0.6\columnwidth]{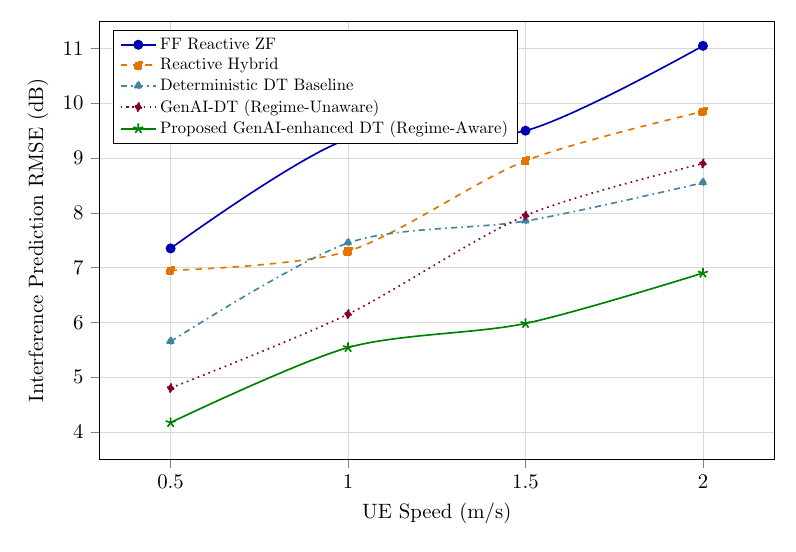}
\caption{
Interference-prediction RMSE versus UE speed in Scene~B.
}
\label{fig:mobility_speed}
\end{figure}

Fig.~\ref{fig:mobility_speed} evaluates mobility sensitivity in the more challenging Scene~B for UE speeds from $0.5$ to $2$ m/s. Prediction error generally increases with speed because faster mobility produces greater channel and position uncertainty over the prediction horizon. The proposed regime-aware GenAI-DT maintains the lowest RMSE across all considered speeds, indicating that its prediction advantage is preserved beyond the nominal $1$ m/s pedestrian setting. The increasing RMSE at higher speeds also highlights the expected tradeoff between mobility and future-state predictability.

\section{Conclusions}
\label{conclusion}

This work presented an integrated GenAI-enhanced DT framework for proactive interference management in hybrid NF/FF XL-MIMO systems. By combining a physics-consistent, site-specific DT with stochastic future-state generation, the proposed framework anticipates channel evolution, geometry-dependent blockage, and NF/FF transitions and incorporates them into regime-aware proactive beamforming. The use of multiple plausible future realizations further improves robustness to prediction uncertainty compared with single-state deterministic approaches. Simulations across open and cluttered indoor environments demonstrated improved interference suppression, SINR, outage performance, and prediction accuracy over reactive, deterministic DT, and regime-unaware GenAI-DT benchmarks. Additional evaluations across different UE speeds, together with computational-complexity and latency analysis, further demonstrated the performance--complexity tradeoff of the proposed framework.

\bibliographystyle{IEEEtran}
\bibliography{Ref}

\begin{IEEEbiography}[{\includegraphics[width=1in,height=1.25in,clip,keepaspectratio]{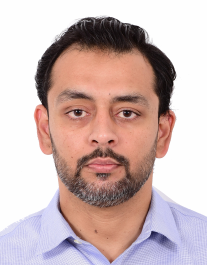}}]{AFAN ALI } (Senior Member, IEEE) Afan Ali received the B.Sc. degree in telecommunications engineering from the National University of Computer and Emerging Sciences (NUCES), Islamabad, Pakistan, in 2008, the M.Sc. degree in engineering from Australian National University (ANU), Canberra, Australia, in 2011, and Ph.D. degree in information and communication engineering from the Northwestern Polytechincal University (NWPU), Xi’an, China, in 2018. From 2009 to 2024, he served at the National Engineering and Scientific Commission (NESCOM), Islamabad, Pakistan, where he progressed through successive research leadership roles. From 2024 to 2025, he was a Postdoctoral Research Fellow in Communications, Signal Processing, and Networking Center (CoSiNC), at the Istanbul Medipol University, Istanbul, Turkey. He received the Best Paper Award 2016 at IEEE Advanced Information Management, Communicates, Electronic and Automation Control Conference (IMCEC), Xi’an, China. He was the recipient of China Outstanding International Student Award 2018, Shaanxi Province, China. He is currently a Postdoctoral Research Fellow at Interdisciplinary Research Center for Communication Systems and Sensing (IRC-CSS), King Fahd University of Petroleum \& Minerals (KFUPM), Saudi Arabia. His research lies at the intersection of wireless communications, active and predictive digital twins, and Generative and Agentic AI for 6G networks. 
\end{IEEEbiography}

\begin{IEEEbiography}[{\includegraphics[width=1in,height=1.25in,clip,keepaspectratio]{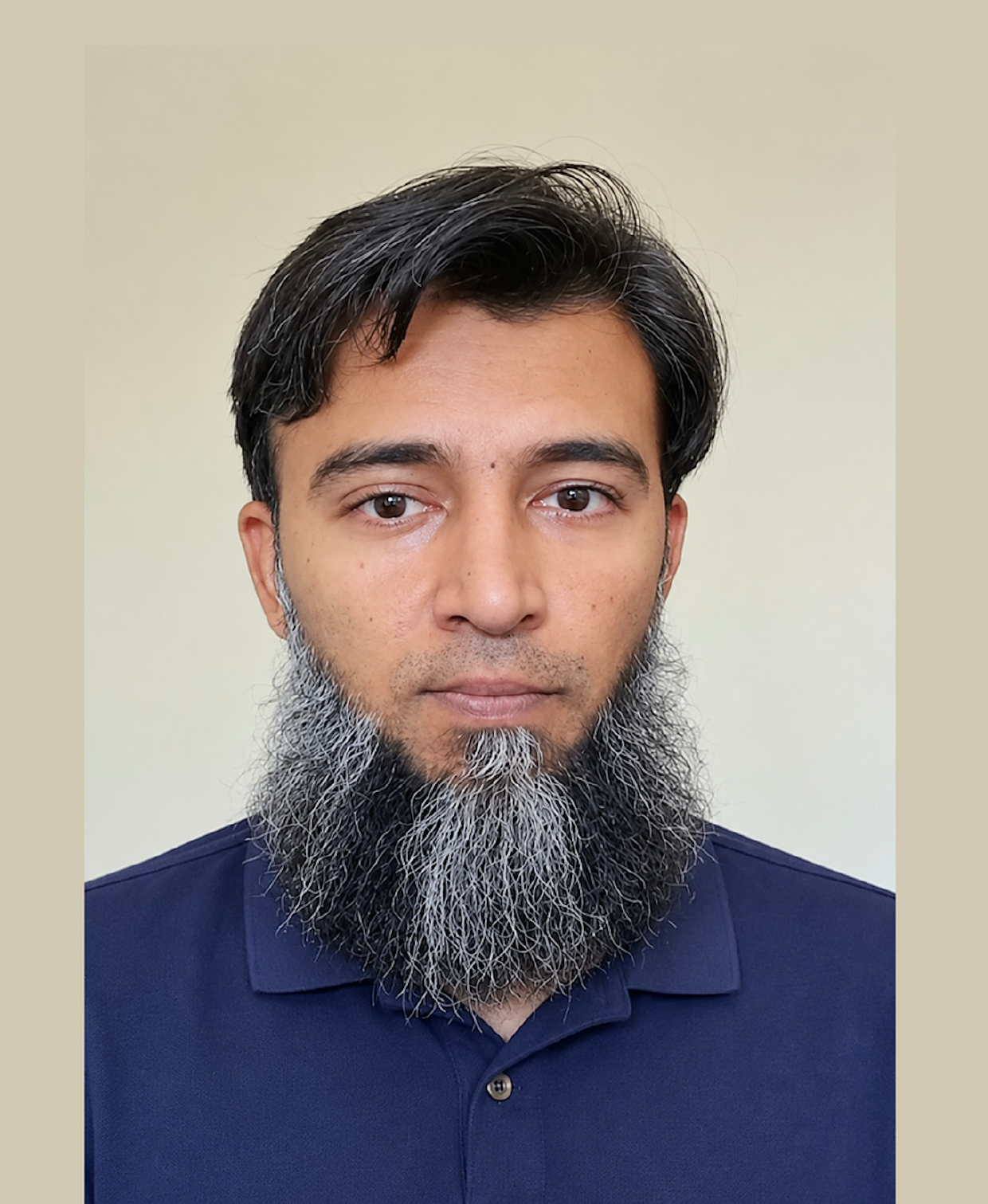}}]{PROF. ALI ARSHAD NASIR } (Senior Member, IEEE) Prof. Ali Arshad Nasir received the Ph.D. degree in telecommunications engineering from the Australian National University, Australia, in 2013, where he worked as a Research Fellow from 2012 to 2015. From 2015 to 2016, he was with the School of Electrical Engineering and Computer Science, National University of Sciences and Technology, Pakistan. He joined the Department of Electrical Engineering, King Fahd University of Petroleum and Minerals, Dhahran, Saudi Arabia, in 2016, where he is currently working as a Professor. His research interests are in the area of signal processing in wireless communication systems. He served as an Editor for IEEE Wireless Communications Letters from 2021 to 2023 and for IEEE Communications Letters from 2024 to 2025. He received the Exemplary Editor Award from IEEE Communications Letters in 2025. He has been serving as an Editor for IEEE Transactions on Communications since 2026.
\end{IEEEbiography}

\begin{IEEEbiography}[{\includegraphics[width=1in,height=1.25in,clip,keepaspectratio]{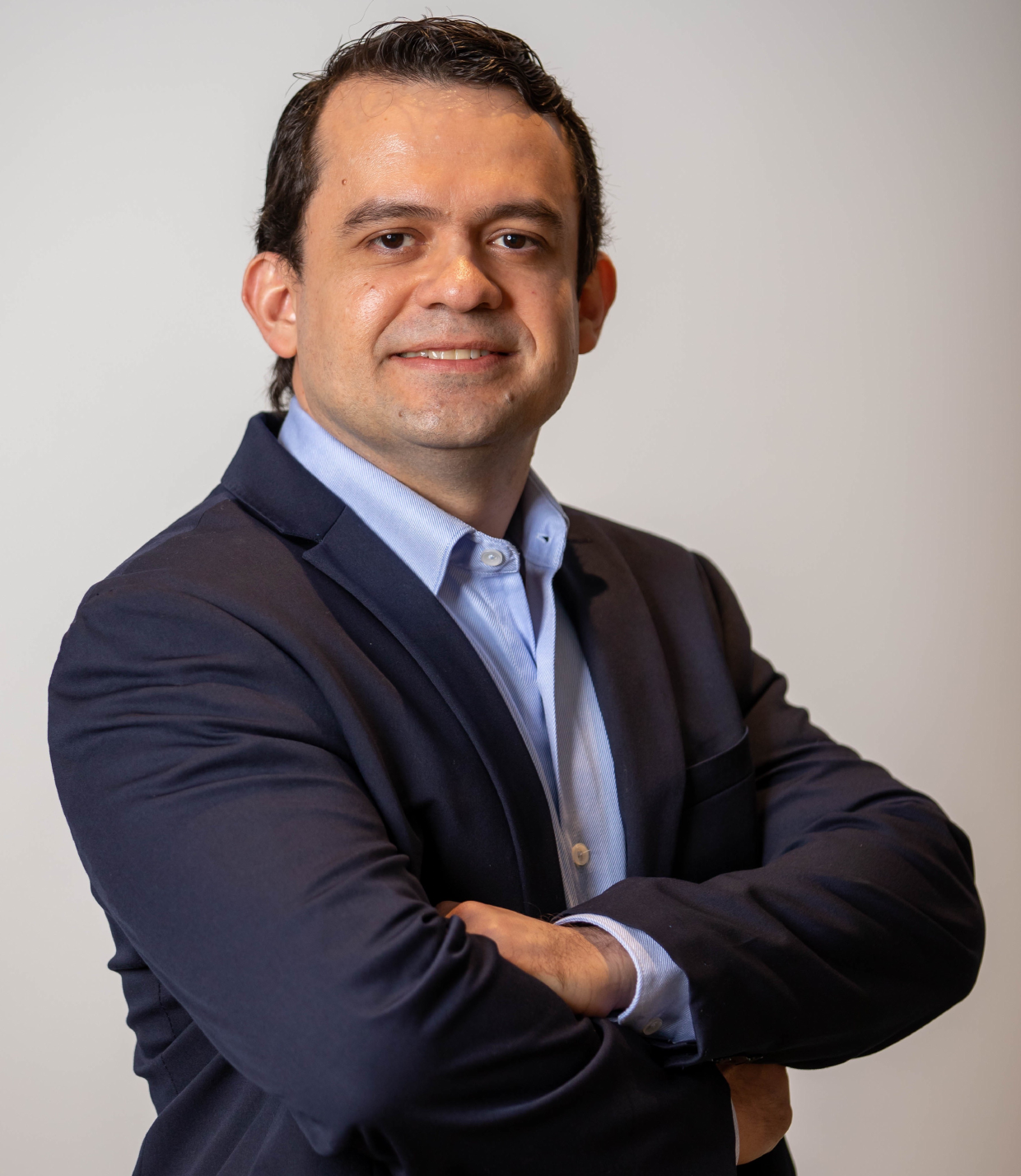}}]{PROF. DANIEL BENEVIDES DA COSTA } (Senior Member, IEEE) Prof. Daniel B. da Costa received the B.Sc. degree in Telecommunications from the Military Institute of Engineering (IME), Rio de Janeiro, Brazil, in 2003, and the M.Sc. and Ph.D. degrees in Electrical Engineering, Area: Telecommunications, from the University of Campinas, SP, Brazil, in 2006 and 2008, respectively. His Ph.D. thesis was awarded the Best Ph.D. Thesis in Electrical Engineering by the Brazilian Ministry of Education (CAPES) at the 2009 CAPES Thesis Contest. From 2010 to 2022, he was Professor at the Federal University of Ceará, Brazil. He also held postdoctoral and visiting positions at INRS-EMT (Canada), Lappeenranta University of Technology (Finland), King Abdullah University of Science and Technology (Saudi Arabia), and Istanbul Medipol University (Turkey), and he was Full Professor at National Yunlin University of Science and Technology (Taiwan). From 2022 to 2024, he was Principal Researcher at the Technology Innovation Institute (TII), Abu Dhabi. He is currently RDIA Chair Professor of Wireless Communications at King Fahd University of Petroleum \& Minerals (KFUPM), Saudi Arabia, and Acting Director of the Interdisciplinary Research Center for Communication Systems and Sensing (IRC-CSS) at KFUPM. Prof. da Costa also serves as Editor-in-Chief of IEEE Communications Letters and Specialty Chief Editor of Frontiers in Communications and Networks, and has held several leadership roles within IEEE societies. 
\end{IEEEbiography}

\end{document}